\documentclass[fleqn,usenatbib]{mnras}

\usepackage{mathptmx}
\usepackage{txfonts}

\usepackage[T1]{fontenc}
\usepackage{ae,aecompl}

\usepackage{graphicx}	

\usepackage{float}

\newcommand{\beq}{\begin{equation}}
\newcommand{\eeq}{\end{equation}}
\newcommand{\beqa}{\begin{eqnarray}}
\newcommand{\eeqa}{\end{eqnarray}}

\newcommand{\mpr}{m_{\rm p}}
\newcommand{\mel}{m_{\rm e}}

\newcommand{\sigmaT}{\sigma_{\rm T}}

\newcommand{\Omegab}{\Omega_{\rm b}}
\newcommand{\Omegam}{\Omega_{\rm m}}
\newcommand{\Omegal}{\Omega_\Lambda}

\newcommand{\Mbh}{M_{\rm{BH}}}
\newcommand{\Msun}{M_\odot}
\newcommand{\Mi}{M_{\rm{i}}}
\newcommand{\mi}{m_{\rm{i}}}
\newcommand{\Mf}{M_{\rm{f}}}
\newcommand{\mf}{m_{\rm{f}}}
\newcommand{\zi}{z_{\rm{i}}}
\newcommand{\zf}{z_{\rm{f}}}
\newcommand{\te}{t_{\rm{S}}}
\newcommand{\Le}{L_{\rm{Edd}}}

\newcommand{\Mdote}{\dot{M}_{\rm{E}}}
\newcommand{\Mdot}{\dot{M}}
\newcommand{\mdot}{\dot{m}}
\newcommand{\mdotav}{\langle\dot{m}\rangle}
\newcommand{\kduty}{k_{\rm{duty}}}

\newcommand{\rin}{r_{\rm{in}}}
\newcommand{\rout}{r_{\rm{out}}}
\newcommand{\Tmax}{T_{\rm{max}}}
\newcommand{\Emax}{E_{\rm{max}}}
\newcommand{\Tcmb}{T_{\rm{CMB}}}
\newcommand{\dTk}{\Delta T_{\rm{K}}}
\newcommand{\Tk}{T_{\rm{K}}}
\newcommand{\Ts}{T_{\rm{s}}}
\newcommand{\Tb}{T_{\rm{b}}}

\newcommand{\fsc}{f_{\rm sc}}
\newcommand{\Fx}{F_{\rm X}}
\newcommand{\Dl}{D_{\rm L}}

\newcommand{\NH}{N_{\rm{H}}}
\newcommand{\Lmean}{\bar{\lambda}}
\newcommand{\Lamean}{\bar{\theta}}
\newcommand{\fheat}{f_{\rm{heat}}}

\newcommand{\nh}{n_{\rm H}}
\newcommand{\nhi}{n_{\rm HI}}
\newcommand{\nhii}{n_{\rm HII}}
\newcommand{\nhe}{n_{\rm He}}
\newcommand{\nhei}{n_{\rm HeI}}
\newcommand{\nheii}{n_{\rm HeII}}
\newcommand{\nheiii}{n_{\rm HeIII}}

\newcommand{\xhi}{x_{\rm HI}}
\newcommand{\xhii}{x_{\rm HII}}
\newcommand{\xhei}{x_{\rm HeI}}
\newcommand{\xheii}{x_{\rm HeII}}
\newcommand{\xheiii}{x_{\rm HeIII}}

\newcommand{\nel}{n_{\rm e}}

\newcommand{\Gammahi}{\Gamma_{\rm HI}}
\newcommand{\Gammahei}{\Gamma_{\rm HeI}}
\newcommand{\Gammaheii}{\Gamma_{\rm HeII}}

\newcommand{\alphahii}{\alpha_{\rm HII}}
\newcommand{\alphaheii}{\alpha_{\rm HeII}}
\newcommand{\alphaheiii}{\alpha_{\rm HeIII}}

\newcommand{\sigmahi}{\sigma_{\rm HI}}
\newcommand{\sigmahei}{\sigma_{\rm HeI}}
\newcommand{\sigmaheii}{\sigma_{\rm HeII}}

\newcommand{\Ihi}{I_{\rm HI}}
\newcommand{\Ihei}{I_{\rm HeI}}
\newcommand{\Iheii}{I_{\rm HeII}}

\newcommand{\nshi}{N_{\rm s,HI}}
\newcommand{\nshei}{N_{\rm s,HeI}}
\newcommand{\rhob}{\rho_{\rm b}}

\newcommand{\rnull}{\theta_0}
\newcommand{\rhalf}{\theta_{1/2}}
\newcommand{\Tbbgr}{T_{\rm{b,bgr}}}
\newcommand{\Fnu}{F_{\nu}}

\title[Universal 21 cm signature of growing massive black holes]{A
  universal 21 cm signature of growing massive black holes in the
  early Universe}  
\author[S. Sazonov and
  I. Khabibullin]{S. Sazonov$^{1,2}$\thanks{E-mail:
    sazonov@iki.rssi.ru} and I. Khabibullin$^{3,1}$\\ 
$^{1}$Space Research Institute, Russian Academy of Sciences,
  Profsoyuznaya 84/32, 117997 Moscow, Russia\\ 
$^{2}$National Research University Higher School of Economics,
  Myasnitskaya ul. 20, 101000 Moscow, Russia\\
$^{3}$Max Planck Institute for Astrophysics,
  Karl-Schwarzschild-Strasse 1, D-85741 Garching, Germany 
}

\begin{document}
\label{firstpage}
\pagerange{\pageref{firstpage}--\pageref{lastpage}}
\maketitle

\begin{abstract}
  
There is a hope that looking into the early Universe with
next-generation telescopes, one will be able to observe the early
accretion growth of supermassive black holes (BHs) when their masses
were $\sim 10^4$--$10^6\Msun$. According to the standard accretion
theory, the bulk of the gravitational potential energy released by 
radiatively efficient accretion of matter onto a BH in this mass 
range is expected to be emitted in the extreme UV--ultrasoft X-ray
bands. We demonstrate that such a 'miniquasar' at $z\sim 15$ should
leave a specific, localized imprint on the 21~cm cosmological
signal. Namely, its position on the sky will be surrounded by a region
with a fairly sharp boundary of several arcmin radius, within which
the 21~cm brightness temperature quickly grows inwards from the
background value of $\sim -250$~mK to $\sim +30$~mK. The size of this
region is only weakly sensitive to the BH mass, so that the flux
density of the excess 21~cm signal is expected to be $\sim
0.1$--0.2~mJy at $z\sim 15$ and should be detectable by the Square
Kilometer Array. We argue that an optimal strategy would be to search
for such signals from high-$z$ miniquasar candidates that can be found
and localized with a next-generation X-ray mission such as {\it
  Lynx}. A detection of the predicted 21 cm signal would provide a
measurement of the growing BH's redshift to within $\Delta
z/(1+z)\lesssim 0.01$.
  
\end{abstract}


\section{Introduction}
\label{s:intro}

The discovery of powerful quasars at $z\approx 7.5$ \citep{banetal18}
implies that fully fledged supermassive black holes (BHs) as heavy as
$\Mbh\sim 10^9\Msun$ already existed when the Universe was just 700
million years old. If a significant fraction of this mass has been
accumulated by accretion at a nearly critical rate, then the growth of
such objects must have started very early on (at $z\gtrsim 20$) from
seeds that already had masses $\sim 10^3\Msun$ or more. What kind of
object these seeds were is one the most interesting open questions in
astrophysics (see \citealt{volonteri10,latfer16,wooetal18} for
reviews). There is a hope that looking into the $z\sim 20$--10 epochs
with next-generation telescopes, one will be able to observe the early
accretion growth of supermassive BHs when their masses were $\sim
10^4$--$10^6\Msun$. Hereafter, we will refer to such accretors as
'miniquasars'. 

One of the most promising ways to find such miniquasars is in X-rays,
since we know that both stellar-mass and supermassive black holes emit
copious amounts of X-rays during accretion (X-ray binaries, XRBs, and
active galactic nuclei, AGN, respectively). Unfortunately, even the
sensitivity of the {\it Chandra} X-ray Observatory is not sufficient
for detecting miniquasars at $z\gtrsim 6$. The situation will change
dramatically if a mission such as the proposed {\it Lynx} is
implemented in the future. {\it Lynx} is planned to achieve a
sensitivity as high as $10^{-19}$~erg~cm$^{-2}$~s$^{-1}$ (0.5--2~keV)
in combination with {\it Chandra}-like (arcsecond) angular resolution
and substantial sky coverage ($\sim 400$~arcmin$^2$) in its deep
extragalactic surveys \citep{lynx18,benetal18}. This implies that
X-ray sources with luminosities as low as a few $10^{41}$~erg~s$^{-1}$
(rest-frame 2--10~keV) will be detectable without confusion at $z\sim
15$. Assuming that hard X-ray emission carries a significant ($\sim
10$\%) fraction of the near-Eddington bolometric luminosity of a
miniquasar, {\it Lynx} will be able to detect accreting BHs with
masses as low as a few $10^4\Msun$ in the early Universe.

The bulk of the gravitational potential energy released during
radiatively efficient accretion onto a BH emerges in the form of
quasi-thermal radiation from the accretion disk \citep{shasun73}, with
the effective waveband shifting from the optical--UV for supermassive
BHs to soft X-rays for stellar-mass BHs, as observed in AGN and XRBs
(in so-called soft/high states for the latter,
e.g. \citealt{gilmer14}). For intermediate-mass BHs, the bulk of the
disk's emission is expected to fall into the far UV--ultrasoft X-ray
band. Therefore, due to cosmological redshift, {\it Lynx} will not be
able to detect this primary emission component, but, as already
mentioned above, it should be able to observe additional, harder
radiation that can arise due to Comptonization of thermal emission
from the disk in its hot corona. In principle, the redshifted thermal
emission from miniquasars could be observed directly in the
optical--infrared band, but since the Eddington luminosity for a BH of
mass $\sim 10^5\Msun$ at $z\sim 15$ corresponds to an AB magnitude of
more than 30, the detection of such miniquasars will be extremely
challenging even with the next-generation IR observatories such as the
{\it James Web Space Telescope} and {\it Wide-field Infrared Survey
  Telescope} (see \citealt{masetal15} for the expected sensitivities
of future surveys with these telescopes).

There is, however, another, indirect way to reveal the primary thermal
radiation from the first miniquasars, which is to observe its impact on
the ambient intergalactic medium (IGM) in the early Universe using the
21~cm spin-flip transition of neutral hydrogen. As has been actively
discussed over the past two decades, the first generations of X-ray
sources could significantly heat the primordial IGM prior to cosmic
reionization and strongly modify the global 21~cm signal from the
$z\sim 15$--10 epochs (see \citealt{priloe12} for a review). A lot of
recent literature on this subject is devoted to discussing the
potentially observable effect of the first generations of stellar-type
X-ray sources and in particular high-mass X-ray binaries (HMXBs,
e.g. \citealt{miretal11,cohetal17,madfra17,sazkha17a}), which likely
were present in significant numbers since the beginning of active star
formation in the Universe \citep{fraetal13}. The bulk of the emission
produced by HMXBs is at energies above 0.5~keV and since such photons
can travel large distances before being photoabsorbed in the IGM, the
main effect of HMXBs is expected to be a global enhancement of the IGM
temperature together with large-scale fluctuations reflecting the
large-scale structure of the early Universe
(e.g. \citealt{prifur07,rosetal17}).

In contrast, miniquasars, which presumably have much softer energy
spectra compared to HMXBs, should mainly heat the IGM in their
relatively close vicinity. One may thus expect such sources to be
surrounded by compact regions of specific 21~cm signal. Although
finding such 21~cm features in a blind search might be difficult even
for the most ambitious upcoming radio interferometers such as the Square
Kilometer Array (SKA, \citealt{meletal13}), such a search could be
greatly faciliated if carried out around miniquasar candidates found
via their coronal X-ray emission with a mission like {\it Lynx}. We
elaborate on this idea below. Before proceeding, we note that there
have been plenty of studies addressing the impact of quasars and
miniquasars on the IGM and the associated 21~cm signal
(e.g. \citealt{madetal04,ricost04,chuetal06,thozar08,yajli14,fiaetal17,ghaetal17,boletal18,vasetal18}),
but the novelty of our study is its focus on the miniquasar's primary,
thermal emission component and the synergy of 21~cm and X-ray
observations. 
 
The following values of cosmological parameters are used throughout
the paper: $\Omegam=0.309$, $\Omegal=1-\Omegam$, $\Omegab=0.049$,
$H_0=68$~km~s$^{-1}$~Mpc$^{-1}$ and $Y=0.246$ (helium mass fraction)
\citep{planck16}.

\section{Model}
\label{s:model}

Suppose a BH has an initial mass $\Mi$ at redshift $\zi$ and accretes
matter until epoch $\zf$, reaching a final mass $\Mf$. If the
accretion proceeded at a critical (Eddington limited) rate $\Mdote$,
the BH mass would be increasing exponentially,
\beq 
M(t)=\Mi e^\frac{t}{\te},
\eeq
on the Salpeter time scale
$\te=\frac{\epsilon}{1-\epsilon}\frac{c^2\Msun}{\Le(\Msun)}$, where
$\Le$ is the Eddington luminosity and $\epsilon$ is the radiation
efficiency. Adopting for simplicity $\epsilon=0.1$ (as is
approximately true for standard accretion disks), $\te\approx 5\times
10^7$~yr. Therefore, the average rate expressed in units of the
critical rate (usually referred to as the Eddington ratio), at which
the BH accretes mass between epochs $\zi$ and $\zf$ is 
\beq
\mdotav\equiv\frac{\Mdot}{\Mdote}=\frac{\te}{t(\zi,\zf)}
\ln\frac{\Mf}{\Mi},
\eeq
where $t(\zi,\zf)$ is the cosmic time between $\zi$ and $\zf$. 

It is unlikely though that the BH will accrete matter at a constant
rate over a cosmologically long period of time. In reality, accretion
onto the BH will be determined by evolving external and
internal (with respect to the host galaxy) conditions and is likely to
be an intermittent process. Therefore, in our simulations,  described
below, we assumed that there are periods of active accretion when the
Eddington ratio takes a fixed value $\mdot$ and passive periods when
$\mdot=0$. We further assume that these two types of intervals
alternate in a random fashion\footnote{We took the duration of these
  intervals to be $\Delta t=10^4$ or $10^5$~yr, with the results being
  insensitive to this choice as long as $\Delta t\ll\te$.}, so that
the duty cycle of BH activity is
\beq
\kduty=\frac{\mdotav}{\mdot}.
\eeq
By definition, $\mdot\ge\mdotav$, and we also assume that $\mdot<1$,
i.e. we do not consider supercritical accretion in this study. 

\subsection{Emission spectrum}
\label{s:spec}

One of the key aspects for this study is the spectrum of the radiation
emitted by the accreting BH. According to the standard accretion
theory \citep{shasun73}, a geometrically thin, optically thick
accretion disk around a BH is characterized by a $\propto r^{-3/4}$
temperature profile (except in the very narrow innermost region, where
only a small fraction of the total luminosity is emitted) and
generates multicolor, nearly blackbody radiation with a spectrum
(specific luminosity as a function of energy)
\beq
L_E(E)\propto \int_{\rin}^{\rout} rB_E(E,T(r))\,dr,
\label{eq:le}
\eeq
where $\rin$ and $\rout$ are the disk's inner and outer radii and
$B_E$ is the Planck function.

The maximum temperature of the disk is
\beq
k\Tmax\approx 1.2\left(\frac{\mdot}{m}\right)^{1/4}\,{\rm keV},
\label{eq:tmax}
\eeq
where $m(t)$ is the growing BH mass expressed in solar
masses. According to the standard theory, this temperature is achieved
at $(49/36) r_0$, where $r_0$ is the radius of the innermost 
stable circular orbit, but a fairly good approximation is that the
disk temperature reaches this value at $\rin$ and then decreases as
$T(r)=\Tmax(r/\rin)^{-3/4}$ at $r>\rin$. For the purposes of this
study it can also be safely assumed that $\rout\to\infty$. The
spectrum given by equation~(\ref{eq:le}) can then be approximated by
the power law $L_E\propto E^{1/3}$ at $E\lesssim 0.3k\Tmax$ and by the
blackbody spectrum with a temperature of $0.7\Tmax$ at $E\gtrsim
2k\Tmax$ \citep{maketal86}, with its maximum (when plotted in units of
$EL_E$) being at $\Emax\approx 2.35k\Tmax$. The normalization constant
in equation~(\ref{eq:le}) is determined by the condition 
\beq
\int L_E\,dE=\mdot\Le (m).
\eeq

The model described above is widely known as a multicolor disk
blackbody model ({\it diskbb} in {\small XSPEC}, \citealt{arnaud96}),
and we have chosen it as our baseline spectral model. This choice is
primarily motivated by the fact that 
the standard accretion disk theory provides a satisfactory description
of the observed spectral energy distributions (SED) of (i) XRBs in
their soft/high states (when $\mdot\gtrsim 0.1$); namely, their
dominant emission component is well described by {\it diskbb} with
$k\Tmax\lesssim 1$~keV, as expected from equation~(\ref{eq:tmax}) for
the stellar masses ($m\lesssim 10$) of the BHs in XRBs (see
\citealt{donetal07} for a review) and (ii) AGN -- supermassive BHs
accreting at $\mdot\sim 0.01$--1, for which the peak of the SED is
observed in the optical-UV (the so-called big blue bump,
e.g. \citealt{elvetal94,teletal02,sazetal04}), again as expected from
equation~(\ref{eq:tmax}) for the high ($m\sim 10^7$--$10^9$) BH masses
of AGN. 

In reality, observations reveal significant deviations of XRB and AGN
spectra from the simple multicolor disk blackbody model described
above, and these deviations can be generally accounted for by the
conditions at the inner boundary of the accretion disk, radiative
transfer effects in the disk's atmosphere and relativistic corrections
(e.g. \citealt{korbla99,meretal00,davetal05,donetal12}). However, in
view of other, larger uncertainties related to the problem in hand (in
particular in the BH mass and accretion rate), we do not take these
subtleties into account. 

Arguing further from analogy with XRBs and AGNs, it is likely that a
miniquasar's emission spectrum has an additional, harder component 
due to the Comptonization of part of the thermal radiation from the
disk in its hot corona. We simulate this plausible situation by
modifying our baseline {\it diskbb} model by the {\it simpl}
\citep{steetal09} model [specifically we use {\it simpl(dikbb)} in
  {\small XSPEC}], which provides a simplified description of
Comptonization by converting a given fraction, $\fsc$, of soft thermal
photons into high-energy ones. Another free parameter of this model is
the photon index, $\Gamma$, of the power-law component. We use
$\Gamma=2$ and $\fsc=0.05$ as fiducial values. The assumed spectral
slope is close to those of observed hard X-ray tails in XRBs and AGN
and is convenient in use since no $k$-correction is then needed in
converting luminosities to fluxes. The adopted $\fsc$ value implies
that the power-law component, if it continues up to $E\sim 100$~keV,
contains $\sim 25$\% (with a very weak dependence on the disk
temperature, i.e. on $m$ and $\mdot$) of the miniquasar's bolometric
luminosity, in overall agreement with observations of XRBs in their
high state (e.g. \citealt{donetal07}) and AGN
(e.g. \citealt{sazetal04}). Note that for the adopted values of the
parameters, $\Gamma$ may be considered the slope of the high-energy
part of the spectrum at $E\gtrsim 10 k\Tmax$, where $\Tmax$ is given
by equation~(\ref{eq:tmax}).

As already mentioned, our current treatment is resticted to the case
of subcritical accretion ($\mdot<1$). In reality, in some miniquasars
and/or at some stage of their evolution accretion may proceed at a
supercritical rate. Consideration of such a case would require
adopting a substantially different spectral model, as suggested by the
measured spectra of individual ultraluminous X-ray sources in nearby
galaxies (e.g. \citealt{sazetal14,kaaetal17}), the collective X-ray
spectrum of such sources in the local Universe \citep{sazkha17b} and
theoretical considerations (e.g. \citealt{naretal17,taketal19}). 
  
\subsection{Intragalactic absorption}
\label{s:abs}

\begin{figure}
\centering
\includegraphics[width=\columnwidth,viewport=20 180 560 720]{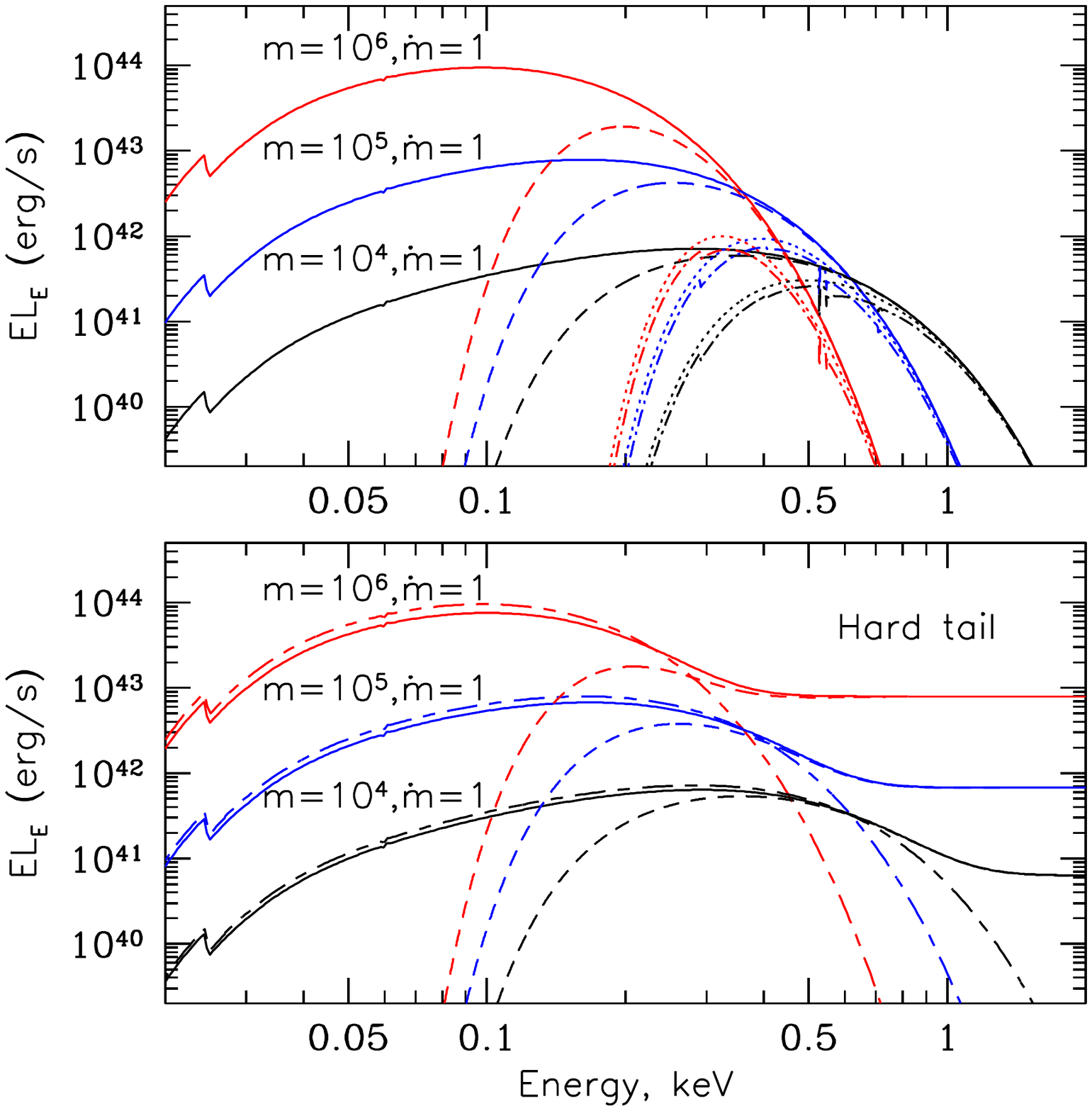}  
\caption{{\it Top panel:} Multicolor disk blackbody emission spectra
  (in the source's rest frame, in units of specific luminosity
  multiplied by photon energy) modified by absorption in the
  miniquasar's host galaxy, for $\mdot=1$, different BH masses,
  $m=10^4$ (black), $10^5$ (blue) and $10^6$ (red), and various
  absorption columns and metallicities ($\NH$ in cm$^{-2}$, $Z$):
  ($10^{18}$, 0) -- solid, ($10^{20}$, 0) -- dashed, ($10^{21}$, 0) --
  dotted, ($10^{21}$, 1) -- dash-dotted. {\it Bottom panel:}
  Multicolor disk blackbody emission spectra modified by
  Comptonization ({\it simpl(diskbb)}), for the same BH masses as
  above (shown with the same colors). The solid and dashed curves
  correspond to $\NH=10^{18}$~cm$^{-2}$ and $\NH=10^{20}$~cm$^{-2}$,
  respectively ($Z=0$). The corresponding pure thermal spectra (for
  $\NH=10^{18}$~cm$^{-2}$) are shown with the short dash--long dashed
  lines for comparison.
}
\label{fig:spectra}
\end{figure}

Before reaching the ambient IGM, the extreme UV-soft X-ray radiation
from the miniquasar may be partially photoabsorbed within its host
galaxy. This may happen (i) in the vicinity of the BH if something
like the AGN obscuring torus is present in miniquasars (in that case,
absorption will take place within a certain solid angle only), and/or
(ii) in the more distant regions of the galaxy. Given our scarce
knowledge about the first galaxies and in particular about the
parsec-scale environment of intermediate-mass BHs they may host, and
also taking into account that the miniquasar's radiation can
significantly ionize the interstellar medium in front of it and
thereby strongly diminish the net absorption effect (see, e.g.,
\citealt{sazkha18}), it is hardly possible to reliably predict the
typical line-of-sight absorption column, $\NH$, for the miniquasars in
the early Universe. We therefore consider it a free
parameter. Similarly, we allow the metallicity of the absorbing gas to
vary from $Z=0$ (pure H--He gas) to $Z=1$ (normal chemical
composition): although the first galaxies (at $z\sim 20$--10) were
likely metal poor, the immediate surroundings of miniquasars might
have been significantly metal enriched because they were probably the
sites of strong star formation activity.

Figure~\ref{fig:spectra} (top panel) shows examples of (rest-frame)
spectra of miniquasars for various values of model parameters, namely
$m$ (assuming $\mdot=1$), $\NH$ and $Z$ (absorption was modeled by
means of the {\it tbvarabs} model in {\small XSPEC}). We see that for
the range of BH masses and accretion rates expected for miniquasars
and in the absence of absorption, the bulk of the accretion disk's
emission is in the extreme UV--very soft X-ray band, at energies
$E\sim 50$--1000~eV. Even a moderate absorption ($\NH\lesssim
10^{20}$~cm$^{-2}$) will cause a strong reduction of the flux below
$\sim 200$~eV. An addition of metals to the absorbing medium (the
$Z=1$ case) will further reduce the flux, but mostly above the oxygen 
absorption edge at $E=536$~eV (note also that the helium absorption
edge at $E=24.6$~eV is clearly seen in the spectra).

The bottom panel of Fig.~\ref{fig:spectra} shows examples of thermal
spectra modified by Comptonization, as described above. For the
adopted value, $\fsc=0.05$, of the fraction of Comptonized photons,
the hard tail starts to dominate over thermal emission at $\sim 1$~keV
for the lowest mass BH ($m=10^4$) and already at $\sim 300$~eV for the
highest mass BH ($m=10^6$).

\section{A crude estimate of the expected heating}
\label{s:estimates}

To a first approximation, the thermal disk emission from miniquasars
(with $m\sim 10^4$--$10^{6}$) in the presence of moderate absorption
($\NH\lesssim 10^{20}$~cm$^{-2}$) may be characterized by a narrow
spectrum around an energy $\sim 300$~eV (see
Fig.~\ref{fig:spectra}). This allows us to derive order-of-magnitude
estimates for the impact of a miniquasar on the IGM before proceeding 
to detailed computations.

The mean free path of soft X-ray photons of energy $E$ in the
primordial (i.e. nearly neutral H--He gas) IGM of the early Universe
can be approximated as follows  \citep{sazsun15}:
\beq
\Lmean\approx
740\left(\frac{1+z}{11}\right)^{-3}\left(\frac{E}{300~{\rm
    eV}}\right)^{3.2}\,{\rm kpc}.
\label{eq:path_kpc}
\eeq
Within this (proper) distance from the source, $1-e^{-1}\approx 63$\%
of photons of energy $E$ will be photoabsorbed (whereas 95\% of
photons will be absorbed within $3\Lmean$). The distance $\Lmean$
corresponds to an angular size
\beq
\Lamean\approx 2.7
\left(\frac{1+z}{11}\right)^{-2}\left(\frac{E}{300~{\rm
    eV}}\right)^{3.2}\,{\rm arcmin},
\label{eq:path_arcmin}
\eeq
on the sky, which is a reasonably good approximation for $z=20$--10
and $E=100$--1000~eV.

The thermal disk emission from a miniquasar can heat the IGM
efficiently only within a few $\Lmean$, since only an exponentially
decreasing fraction of the miniquasar's luminosity reaches larger
distances. This allows us to readily estimate the expected IGM
temperature increment. The total energy released by the BH during its
growth to mass $M$ is $W=\epsilon Mc^2$, and we may assume that most of
this energy is radiated away over a time of order the Salpeter time
($\te$) just before the epoch when the miniquasar and the associated
21~cm signal are observed, so that, to a first approximation,
we can ignore any effects associated with the expansion of the
Universe. We may further assume that all of this energy has been
absorbed within a volume of radius $\sim\Lmean$ (the corresponding
light travel time proves to be shorter than $\te$). Assuming that the
miniquasar ionizes the surrounding medium only moderately (i.e. the
ionization degree of hydrogen is less than a few per cent, which is a
good approximation for the bulk of the affected volume), we may
roughly estimate the mean fraction of the energy of soft X-ray photons
that goes into heating the IGM as $\fheat\sim 0.2$
\citep{fursto10}. Taking into account that the hydrogen space density
changes with redshift as $\nh(z)\approx 2.6\times
10^{-4}[(1+z)/11]^3$~cm$^{-3}$, we can write
\beq
\fheat W=\frac{4\pi\Lmean^3}{3}\frac{3}{2}\nh k\dTk,
\eeq
where $k$ is the Boltzmann constant, and finally determine the
expected IGM temperature increment assuming $E=300$~eV:
\beq
\dTk\sim
100\frac{\fheat}{0.2}\frac{\epsilon}{0.1}\left(\frac{1+z}{11}\right)^{6}\frac{m}{10^4}\,{\rm K}. 
\label{eq:deltat}
\eeq

Comparing this with the cosmic microwave background (CMB) temperature,
$\Tcmb(z)\approx 30[(1+z)/11]$, we come to the conclusion that BHs
with $M\gtrsim 10^4\Msun$ accreting at a nearly critical rate in the
early Universe will be surrounded by well-defined zones
with a radius of a few arcmin within which $\Tk\gtrsim \Tcmb$, and
these regions are thus expected to be 21~cm emitters, in contrast to
the surrounding sky, which is likely to exhibit 21~cm absorption at
$z\gtrsim 10$. Importantly, for $m\gtrsim 10^4$, the size of the
heating region is determined simply by the mean free path of soft
X-ray photons, rather than by the radiative power of the
miniquasar. Following the same argument, we may expect that for BHs of
smaller mass, $m\lesssim 10^4$, the region of strong heating
[$\dTk\gtrsim\Tcmb(z)$] will be smaller than $\sim\Lmean$ and its
actual size will be determined by the total energy released
by accretion onto the BH.

\section{Simulations}
\label{s:simul}

Based on the assumptions outlined in \S\ref{s:model}, we performed a
series of numerical calculations of IGM heating and associated 21~cm
emission/absorption in the vicinity of miniquasars in the early
Universe. We limited simulations to a redshift range of $z=20$--10
and neglected any global heating (i.e. outside the region affected by
the miniquasar) of the IGM by X-ray sources and/or other
mechanisms\footnote{In particular, by low-energy cosmic rays from the
  first supernovae \citep{sazsun15,leietal17}.}. We adopted the
following initial parameters of the IGM:
$\xhii\equiv\nhii/\nh=2.2\times 10^{-4}$ (hydrogen ionization
fraction), $\xheii\equiv\nheii/\nhe=0$, $\xheiii\equiv\nheiii/\nhe=0$
(helium ionization fractions) and either $\Tk=9.3$~K (for $\zi=20$) or
$\Tk=5.4$~K (for $\zi=15$). These were found using {\small RECFAST}
\citep{seaetal99} and correspond to the conditions after cosmic
recombination and adiabatic cooling of the primordial gas. Our
assumption about the absence of significant global heating might be a
good approximation at least at $z\gtrsim 15$, as suggested by the
recent detection of a strong, sky-averaged 21~cm absorption signal in
the Experiment to Detect the Global Epoch of Reionization Signature
(EDGES, \citealt{bowetal18}).

X-ray ionization and heating of the IGM was calculated in
logarithmically binned spherical shells around the miniquasar, out to
a comoving distance of 500~Mpc. Although this maximal distance is
fairly large, we ignored photon travel time effects (i.e. the response
of the IGM to radiation emitted by the central source was considered
instantaneous), since X-ray heating proves to be noticeable only
within $\sim 30$~cMpc of the miniquasar and the corresponding
light travel time at $z\gtrsim 10$ is much shorter than the Salpeter
timescale on which BH growth occurs.

The evolution of the ionization state of hydrogen and helium with time
in a given shell was calculated as follows
[equations~(\ref{eq:dxh})--(\ref{eq:heatrate}) below are adopted from
  \citealt{madfra17,sazkha17a}]: 
\beqa
\frac{d\xhi}{dt} &=& -\xhi\Gammahi+\nel (1-\xhi)\alphahii,\nonumber\\
\frac{d\xhei}{dt} &=& -\xhei\Gammahei+\nel\xheii\alphaheii,\nonumber\\
\frac{d\xheii}{dt} &=&
-\xheii\Gammaheii+\nel\xheiii\alphaheiii-\frac{d\xhei}{dt},\nonumber\\
&&
\label{eq:dxh}
\eeqa
where $\nel$ is the number density of free electrons, $\alphahii$,
$\alphaheii$ and $\alphaheiii$ are the recombination coefficients
(adopted from \citealt{theetal98}), and $\Gammahi$, $\Gammahei$ and
$\Gammaheii$ are the photoionization coefficients, which were
calculated as follows:
\beqa
\Gammahi=\frac{1}{4\pi r^2}\left(\int_{\Ihi}^\infty\frac{L_E
    e^{-\tau(r,E)}}{E}\sigmahi(1+N_{\rm s,HI}(E,\Ihi))dE\right. \nonumber\\ 
  \left.+\int_{\Ihei}^\infty\frac{L_E
    e^{-\tau(r,E)}}{E}\sigmahei\frac{\nhei}{\nhi}N_{\rm
    s,HI}(E,\Ihei)dE\right.\nonumber\\
  \left.+\int_{\Iheii}^\infty\frac{L_E e^{-\tau(r,E)}}{E}\sigmaheii\frac{\nheii}{\nhi}N_{\rm s,HI}(E,\Iheii)dE\right),\nonumber\\
\Gammahei=\frac{1}{4\pi r^2}\left(\int_{\Ihi}^\infty\frac{L_E
    e^{-\tau(r,E)}}{E}\sigmahi\frac{\nhi}{\nhei}N_{\rm s,HeI}(E,\Ihi)dE\right.\nonumber\\
  \left.+\int_{\Ihei}^\infty\frac{L_E e^{-\tau(r,E)}}{E}\sigmahei(1+N_{\rm
    s,HeI}(E,\Ihei))dE\right.\nonumber\\  
  \left.+\int_{\Iheii}^\infty\frac{L_E
    e^{-\tau(r,E)}}{E}\sigmaheii\frac{\nheii}{\nhei}N_{\rm
    s,HeI}(E,\Iheii)dE\right),\nonumber\\ 
\Gammaheii=\frac{1}{4\pi r^2}\int_{\Iheii}^\infty\frac{L_E
  e^{-\tau(r,E)}}{E}\sigmaheii dE,
\label{eq:gammas} 
\eeqa 
where $\Ihi=13.6$~eV, $\Ihei=24.6$~eV and $\Iheii=54.4$~eV are the
ionization thresholds for HI, HeI and HeII, $\nshi$ and $\nshei$ are
the mean numbers of secondary ionizations of HI and HeI (secondary
ionization of HeII is practically unimportant) caused by the fast
photoelectron, with the notation $N_{\rm s,HI}(E,\Ihi)$ meaning that
$N_{\rm s,HI}$ is a function of the photoelectron energy, $E-\Ihi$
(the corresponding dependencies for HI and HeI are adopted from
\citealt{fursto10}), and   
\beq
\tau(r,E)=\int_0^r[\nhi(r')\sigmahi(E)+\nhei(r')\sigmahei(E)+\nheii(r')\sigmaheii(E)]dr'
\eeq
is the IGM photoionization optical depth within radius $r$ from the
source, with the cross-sections $\sigmahi(E)$, $\sigmahei(E)$ and
$\sigmaheii(E)$ adopted from \cite{veretal96}. 

The evolution of the gas temperature with time in a given shell is
given by 
\beq
\frac{d\Tk}{dt}=-2H\Tk+\frac{\Tk}{\mu}\frac{d\mu}{dt}+\frac{2\mu\mpr}{3k\rhob}(\mathcal{H}-\Lambda),
\label{eq:dtk}
\eeq
where the photoionization heating rate is given by
\beqa
\mathcal{H} = \frac{1}{4\pi r^2}\left(\int_{\Ihi}^\infty\frac{L_E
    e^{-\tau(r,E)}}{E}(E-\Ihi)\nhi\sigmahi f_{\rm
  heat}(E,\Ihi)dE\right.\nonumber\\
\left.+\int_{\Ihei}^\infty\frac{L_E e^{-\tau(r,E)}}{E}(E-\Ihei)\nhei\sigmahei
f_{\rm heat}(E,\Ihei)dE\right.\nonumber\\  
\left.+\int_{\Iheii}^\infty\frac{L_E e^{-\tau(r,E)}}{E}(E-\Iheii)\nheii\sigmaheii
f_{\rm heat}(E,\Iheii)dE\right),\nonumber\\
\label{eq:heatrate}
\eeqa
with $H(z)$ being the Hubble constant, $\rhob$ the average baryonic
density of the Universe, $\mu$ the mean molecular weight and $\fheat$
the fraction of the photoelectron energy that goes into gas heating,
which depends on the photoelectron energy as given by \cite{fursto10}.  

The term proportional to $\Lambda$ in equation~(\ref{eq:dtk}) accounts
for the radiative losses arising from collisional and recombinations
processes (the corresponding rates were adopted from
\citealt{theetal98}), as well as for Compton cooling caused by
scattering of the CMB on free electrons\footnote{Inverse Compton
  heating due to the X-ray radiation is negligible except very close
  to the miniquasar.}), which proceeds on the time scale
\beqa  
t_{{\rm CMB}} &=& \frac{3\mel c^2 (n/\nel)}{32\sigmaT\sigma\Tcmb^4(z)}\nonumber\\
&\approx & 8\times
10^{7}\left(\frac{\nel}{n}\right)^{-1}\left(\frac{1+z}{11}\right)^{-4}\,{\rm
  yr},
\label{eq:cmbtime}
\eeqa
where $n$ is the total particle number density, $\sigmaT$ is the
Thomson scattering cross-section and $\sigma$ is the Stefan--Boltzmann
constant, cooling due to collisional processes (including
bremsstrahlung) and CMB scattering proves to be important in the
vicinity of the miniquasar where the gas becomes strongly ionized and
its temperature rises to $\Tk\gtrsim 10^4$~K. However, the cooling
processes typically have a negligible effect on the average parameters
of the IGM heating zone produced by the miniquasar. 

We further assume that the spin temperature, $\Ts$, characterizing the
21~cm transition is everywhere equal to the gas kinetic temperature,
$\Tk$. The EDGES measurement \citep{bowetal18} suggests that it is
indeed the case at $z\lesssim 20$, implying that by that time the
first stars had already created a significant UV (10.2--13.6~eV)
background for decoupling the spin temperature from that of the CMB and
bringing it close to the gas kinetic temperature via the
Wouthuysen--Field effect \citep{wouthuysen52,field58}. Furthemore, the
photoionization of the IGM by soft X-rays from a miniquasar will be
accompanied by the creation of Ly$\alpha$ photons that will further
strengthen the Wouthuysen--Field effect wherever the gas temperature
increases by $\gtrsim 10^3H(z)\te\gtrsim 100$~K on the BH growth
timescale \citep{chuetal06,chemir08}. Under these assumptions and for
the adopted cosmological parameters, the brightness temperature of the
21~cm line is expected to be  
\beq
\Tb=29\xhi\left(\frac{1+z}{11}\right)^{1/2}\left(1-\frac{1+z}{11}\frac{30}{\Tk}\right)\,{\rm
  mK}. 
\label{eq:tb}
\eeq

\section{Results}
\label{s:result}

Our model has the following parameters: the initial and final
redshifts ($\zi$ and $\zf$), the initial and final masses of the BH
($\mi$ and $\mf$, in solar masses), the Eddington ratio during active
accretion phases ($\mdot$), and the intragalactic absorption column
density and metallicity ($\NH$ and $Z$, respectively). We now present
a summary of results obtained for various sets of the parameter
values. Most of the results presented below have been obtained for the
case of purely thermal accretion disk emission, with the spectral
shape as shown in the top panel of Fig.~\ref{fig:spectra}. We specify
explicitly whenever we also take the possible contribution of a
high-energy Comptonization component into account.

\subsection{Gas temperature and 21~cm brightness temperature radial profiles}
\label{s:profiles}

\begin{figure}
\centering
\includegraphics[width=\columnwidth,viewport=20 180 560 720]{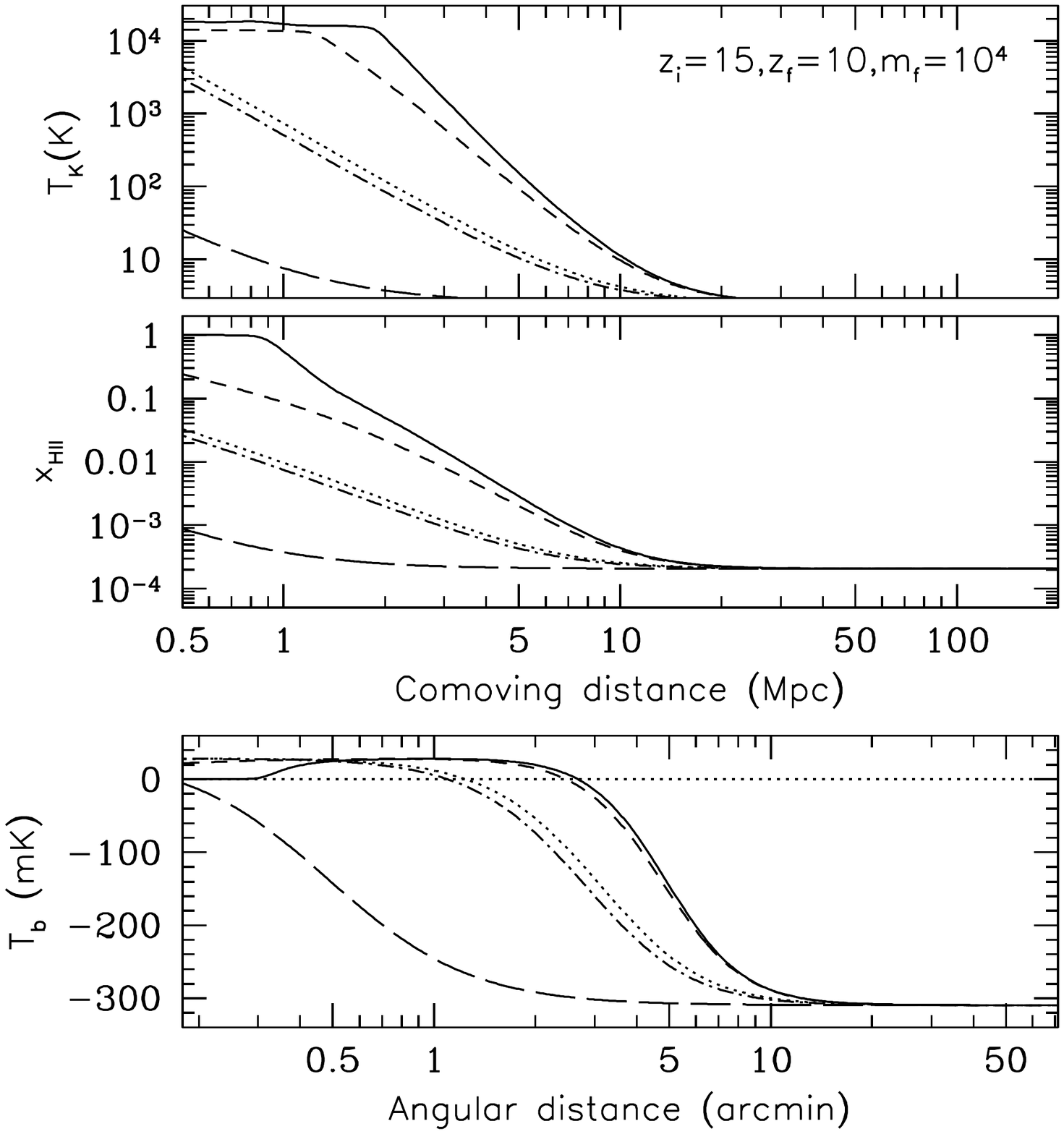}  
\caption{{\it Top panel:} IGM temperature as a function of comoving
  distance from the miniquasar for $\zi=15$, $\zf=10$, $\mi=2\times
  10^3$, $\mf=10^4$, $\mdot=1$ and various parameters of intragalactic
  absorption [$\NH$~(cm$^{-2}$), $Z$]: ($10^{18}$, 0) -- solid,
  ($10^{20}$, 0) -- short-dashed, ($10^{21}$, 0) -- dotted,
  ($10^{22}$, 0) -- long-dashed and ($10^{21}$, 1) --
  dash-dotted. {\it Middle panel:} hydrogen ionization fraction. {\it
    Bottom panel:} 21~cm brightness temperature as a function of
  angular distance from the miniquasar. These plots correspond to
  $\zf$. 
}
\label{fig:profiles_nh_m1e4}
\end{figure}

\begin{figure}
\centering
\includegraphics[width=\columnwidth,viewport=20 180 560 720]{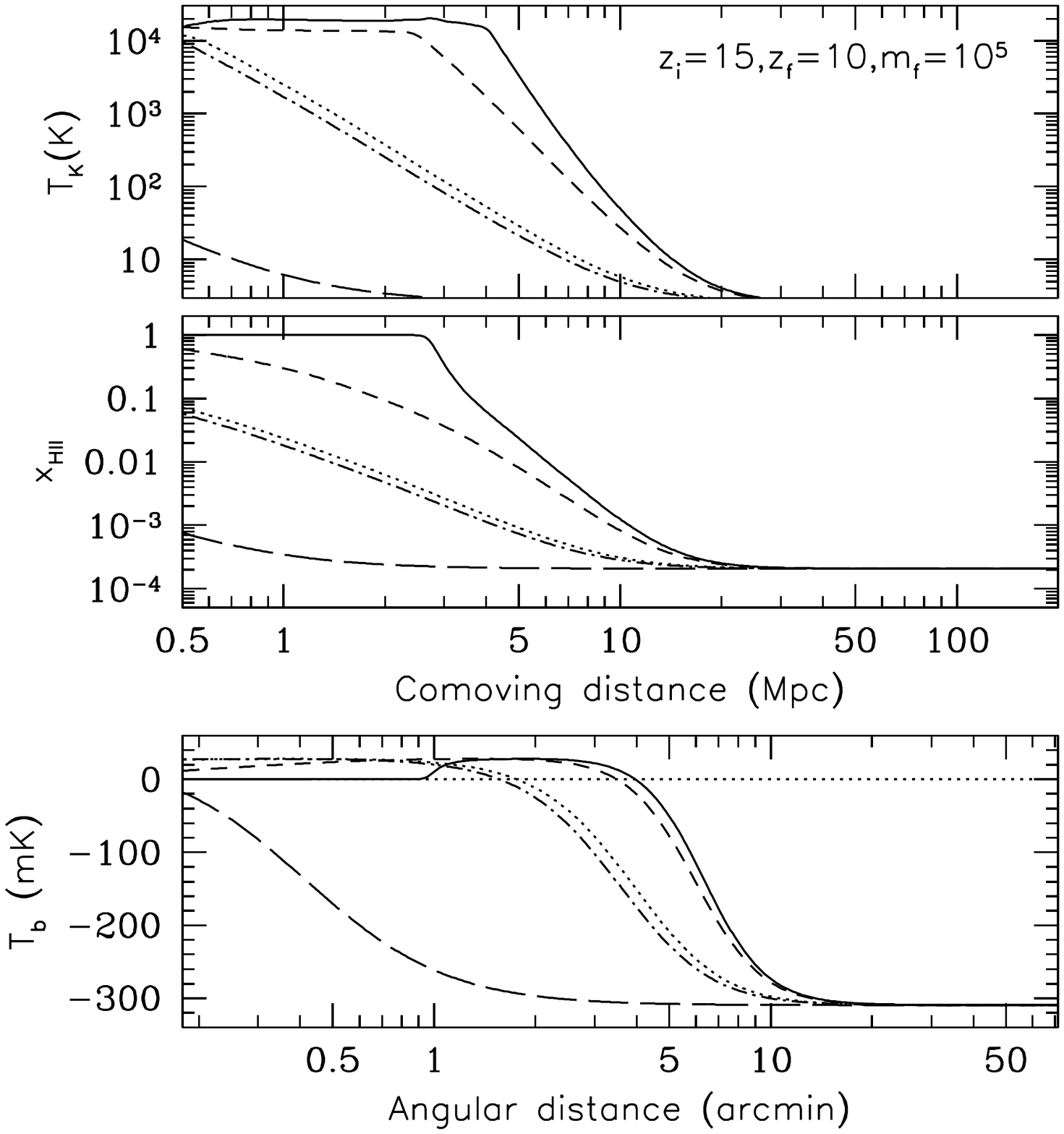}  
\caption{As Fig.~\ref{fig:profiles_nh_m1e4}, but for $\mi=2\times
  10^4$ and $\mf=10^5$. 
}
\label{fig:profiles_nh_m1e5}
\end{figure}

Figure~\ref{fig:profiles_nh_m1e4} shows the IGM temperature and
hydrogen ionization fraction as functions of comoving distance from
the miniquasar and the brightness temperature of the resulting 21~cm
signal as a function of the angular distance in the plane of the sky
for $\zi=15$, $\zf=10$, $\mi=2\times 10^3$, $\mf=10^4$, $\mdot=1$ (the
corresponding accretion duty cycle $\kduty=40$\%) and various
absorption characteristics: $Z=0$, $\NH=10^{18}$, $10^{20}$,
$10^{21}$, $10^{22}$~cm$^{-2}$ and $Z=1$, $\NH=10^{21}$. We see that
the absorption of soft X-rays within the host galaxy leads to less
efficient IGM heating if $\NH\gtrsim 10^{20}$~cm$^{-2}$ and that the
presence of heavy elements ($Z=1$ vs. $Z=0$) proves to be of minor
importance. Therefore, we hereafter focus on the metal-free case,
unless specifically noted otherwise.

Figure~\ref{fig:profiles_nh_m1e5} is analogous to
Fig.~\ref{fig:profiles_nh_m1e4}, but the BH mass has been increased by
an order of magnitude from $\mi=2\times 10^3$, $\mf=10^4$ to
$\mi=2\times 10^4$, $\mf=10^5$. We see that the influence of
intragalactic absorption is similar to the previous case and that the
IGM heating zone has somewhat spread outwards.

\begin{figure}
\centering
\includegraphics[width=\columnwidth,viewport=20 180 560 720]{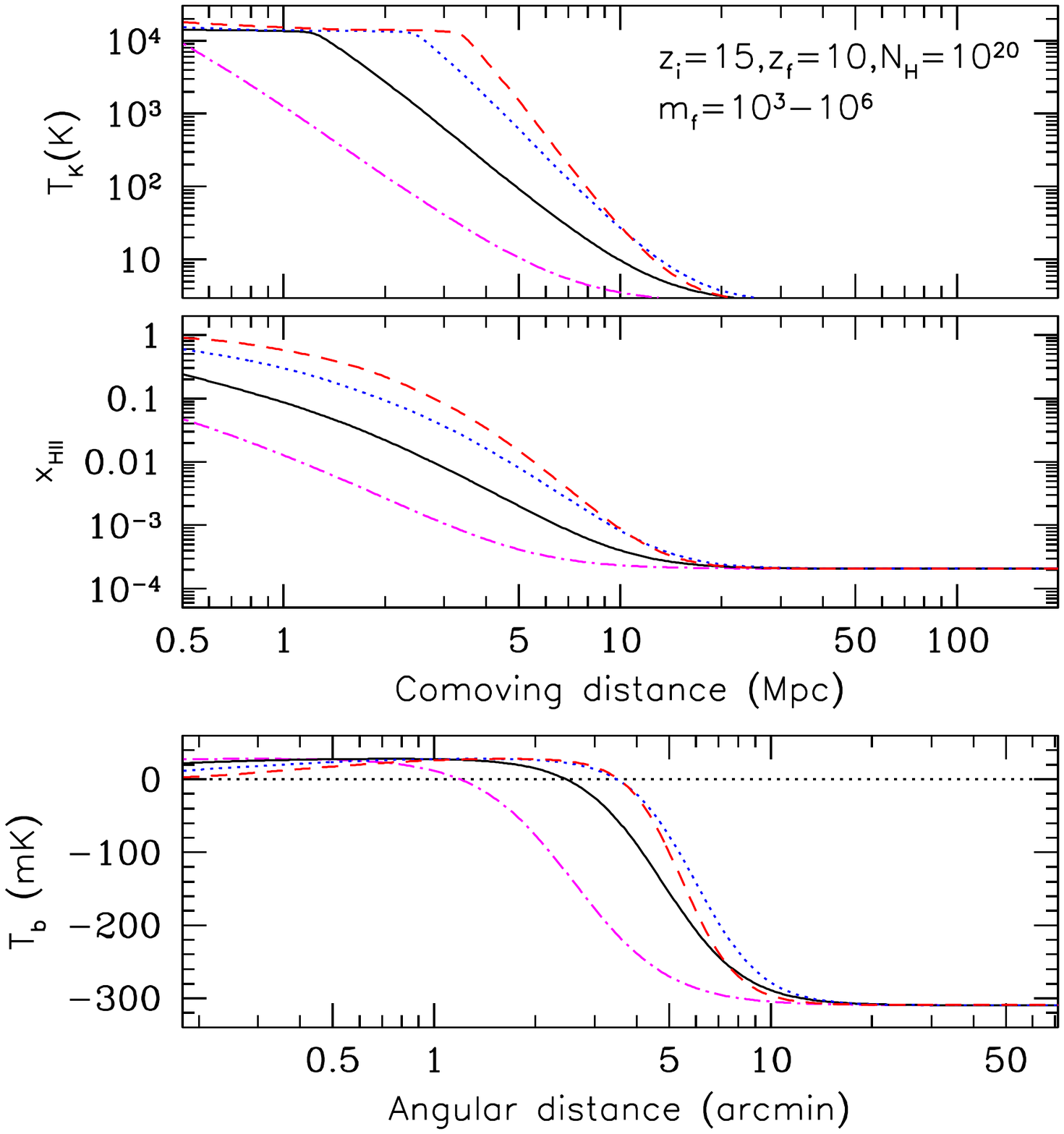}  
\caption{Similar to Fig.~\ref{fig:profiles_nh_m1e4}, but for a fixed
  absorption column ($\NH=10^{20}$~cm$^{-2}$) and various BH masses
  ($\mi$, $\mf$): ($2\times 10^2$, $10^3$) -- dash-dotted magenta,
  ($2\times 10^3$, $10^4$) -- solid black, ($2\times 10^4$, $10^5$) --
  dotted blue, ($2\times   10^5$, $10^6$) -- dashed red. 
}
\label{fig:profiles_mass}
\end{figure}

\begin{figure}
\centering
\includegraphics[width=\columnwidth,viewport=20 180 560 720]{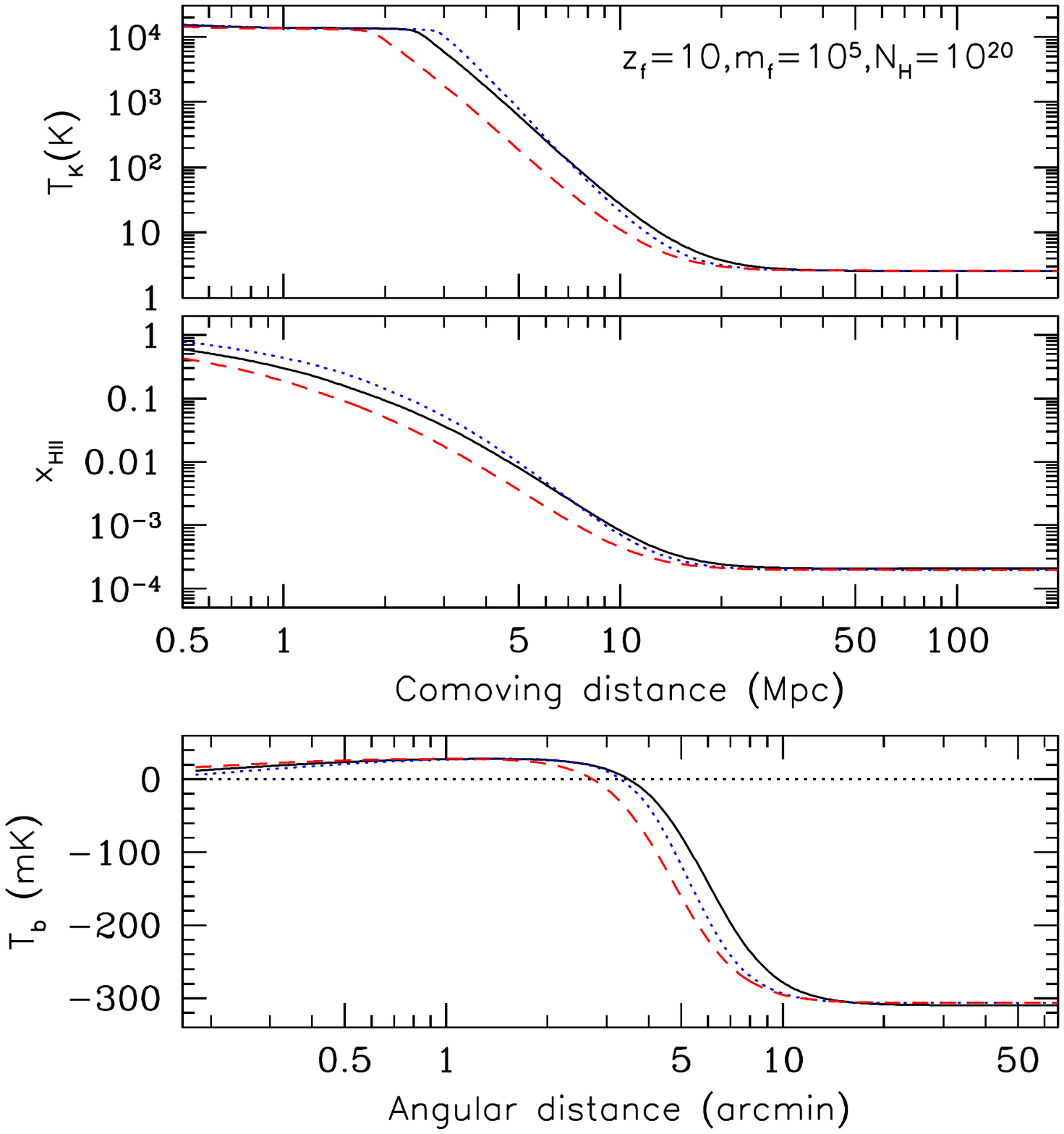}  
\caption{Similar to Fig.~\ref{fig:profiles_nh_m1e4}, but for fixed BH
  mass ($\mi=2\times 10^4$, $\mf=10^5$) and absorption column
  ($\NH=10^{20}$~cm$^{-2}$) and various scenarios of BH growth ($\zi$,
  $\zf$, $\mdot$): (15, 10, 1) -- solid black, (20, 10, 1) -- dotted
  blue, (20, 10, 0.5) -- dashed red.
}
\label{fig:profiles_regime}
\end{figure}

Figure~\ref{fig:profiles_mass} demonstrates the dependence of the
results on the BH mass. Here, we adopted $\zi=15$, $\zf=10$, $\mdot=1$
and $\NH=10^{20}$ and sampled BH masses ($\mi$, $\mf$) from
  ($2\times 10^2$, $10^3$) to ($2\times 10^5$, $10^6$). We see that
although more massive BHs produce much stronger ionization very close
to the source, this has a fairly small effect on the resulting 21~cm
signal, because the innermost region of strong heating is
characterized by a nearly saturated, positive 21~cm brightness
temperature [because $\Tk\gg\Tcmb$, see eq.~(\ref{eq:tb})]. More
important from an observational point of view is what happens at
larger distances where the 21~cm signal changes from emission to
absorption, and we see that the effective size of this region first
noticeably increases on going from $\mf=10^3$ to $\mf=10^4$ and then
remains nearly the same for $\mf=10^5$ and $\mf\le 10^6$ (in fact,
this region is somewhat smaller for $\mf\le 10^6$ than for $\mf=10^5$
because of the smaller number of soft X-ray photons with $E\gtrsim
300$~eV, capable of propagating to large distances, in the former case
-- see Fig.~\ref{fig:spectra}). This behavior is broadly consistent
with the prediction made in \S\ref{s:estimates} that the 21~cm zones
around miniquasars should be largely determined by the total accretion
energy for BHs with $M\lesssim 10^4\Msun$ and by the characteristic
mean free path of accretion disk photons for more massive BHs.

Figure~\ref{fig:profiles_regime} demonstrates the influence of a
particular history of BH growth on the results. Here, we fixed the
final redshift at $\zf=10$, the initial and final BH masses at
$\mi=2\times 10^4$ and $10^5$, respectively, and the absorption column
at $\NH=10^{20}$, and considered three scenarions: (i) $\zi=15$,
$\mdot=1$ (the duty cycle $\kduty=40$\%), (ii) $\zi=20$, $\mdot=1$
($\kduty=27$\%) and (iii) $\zi=20$, $\mdot=0.5$ ($\kduty=55$\%). We
see that the differences in the corresponding $\Tk$ and $\Tb$ profiles
are small.

\begin{figure}
\centering
\includegraphics[width=\columnwidth,viewport=20 180 560 720]{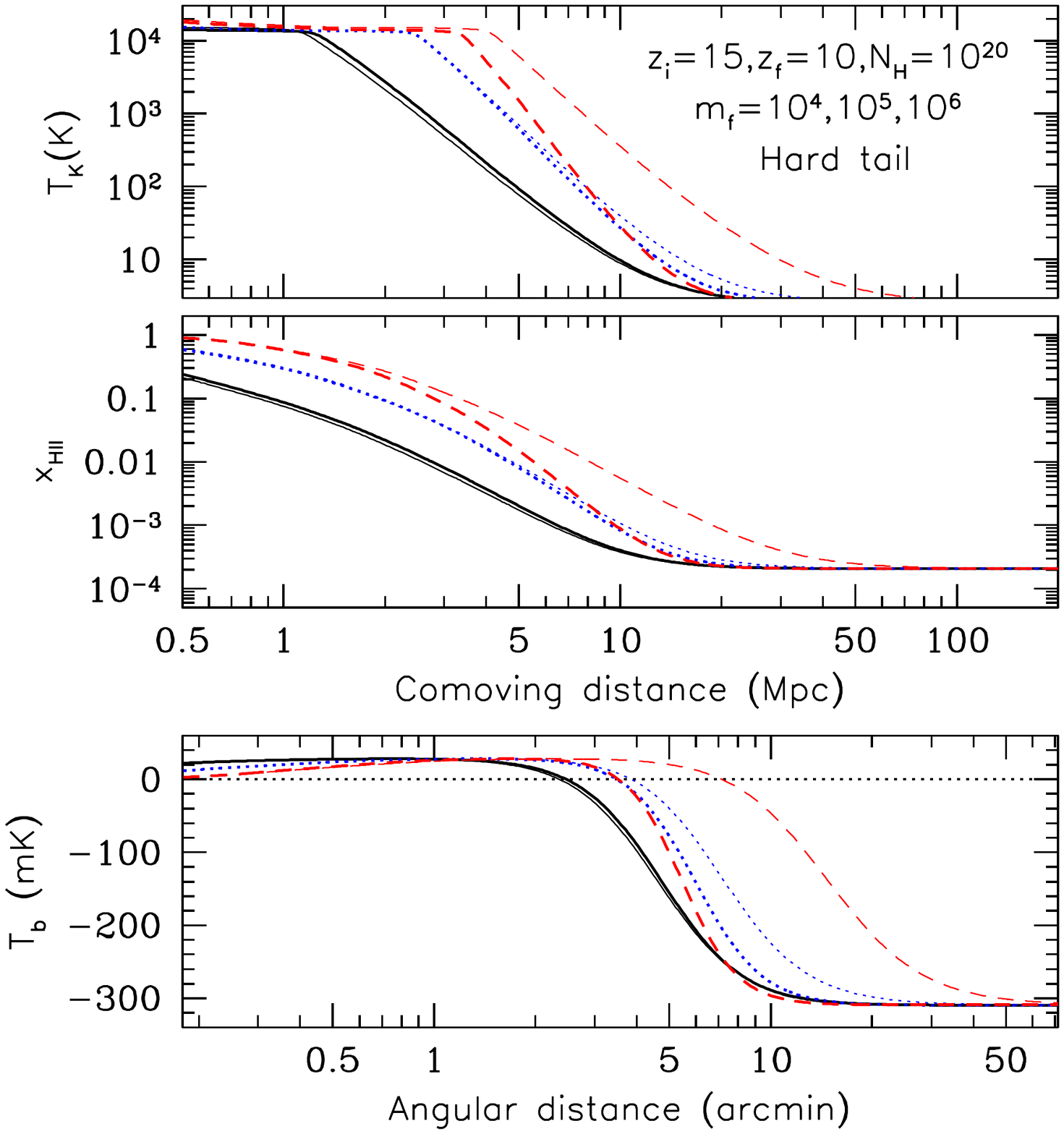}  
\caption{Similar to Fig.~\ref{fig:profiles_nh_m1e4}, for a fixed
  absorption column ($\NH=10^{20}$~cm$^{-2}$, $Z=0$), three sets of BH
  masses ($\mi=2\times 10^3$, $\mf=10^4$ -- solid black, $\mi=2\times
  10^4$, $\mf=10^5$ -- dotted blue, $\mi=2\times 10^5$, $\mf=10^6$ --
  dashed red) and two spectral models: pure thermal disk emission
  ({\it diskbb} -- thick lines) and thermal disk emission with a
  Comptonization tail ({\it simpl}*{\it diskbb} -- thin lines).
}
\label{fig:profiles_simpl}
\end{figure}

So far we have assumed that the incident radiation spectrum is that of
a multicolor accretion disk modified by line-of-sight absorption, as
shown in the top panel of Fig.~\ref{fig:spectra}. We now wish to
investigate the possible effect of an additional hard, power-law
spectral component that may arise due to the Comptonization of soft
photons from the BH accretion disk in its hot corona. To this end, we
carried out calculations for our {\it simpl}*{\it diskbb} spectral
models for $\mf=10^4$, $10^5$ and $10^6$ and $\NH=10^{20}$~cm$^{-2}$,
shown in the bottom panel of Fig.~\ref{fig:spectra}. The
resulting $\Tk$, $\xhii$ and $\Tb$ radial profiles are compared in
Fig.~\ref{fig:profiles_simpl} with those computed without the hard
X-ray component. We see that the 21~cm zone is almost unaffected by
the hard spectral component for the least massive BH ($\mf=10^4$),
somewhat broadens in the intermediate mass case ($m=10^5$), and
becomes substantially (by a factor of $\sim 2$) larger for the
heaviest BH ($m=10^6$). The last result is unsurprising, because the
corresponding X-ray spectrum (see Fig.~\ref{fig:spectra}) is
dominated by the power-law component already at $E\sim 300$~eV
(partially because of the adopted substantial line-of-sight absorption
of $10^{20}$~cm$^{-2}$).

\subsection{Characteristic size of the 21~cm zone}
\label{s:radii}

From the above comparison of the computed $\Tb$ radial profiles a
preliminary conclusion may be drawn that the spatial extent of the
21~cm signal associated with a high-redshift miniquasar will only
weakly depend on the properties of the latter. For more quantitative
assessment, we define two characteristic angular sizes: $\rnull$ --
the projected distance from the miniquasar at which the 21~cm signal
changes from emission to absorption, i.e. $\Tb(\rnull)=0$, and
$\rhalf$ -- the radius at which the brightness temperature of the
absorption signal is half the 'background' value (the 21~cm brightness
temperature outside of the miniquasar heating zone),
i.e. $\Tb(\rhalf)=\Tbbgr/2$. Under our assumptions that there is no
global IGM heating and that the 21~cm spin temperature is coupled to
the IGM kinetic temperature, $\Tbbgr=-245$~mK and $-307$~mK at
$\zf=15$ and $\zf=10$, respectively. 

\begin{figure}
\centering
\includegraphics[width=\columnwidth,viewport=20 180 560 720]{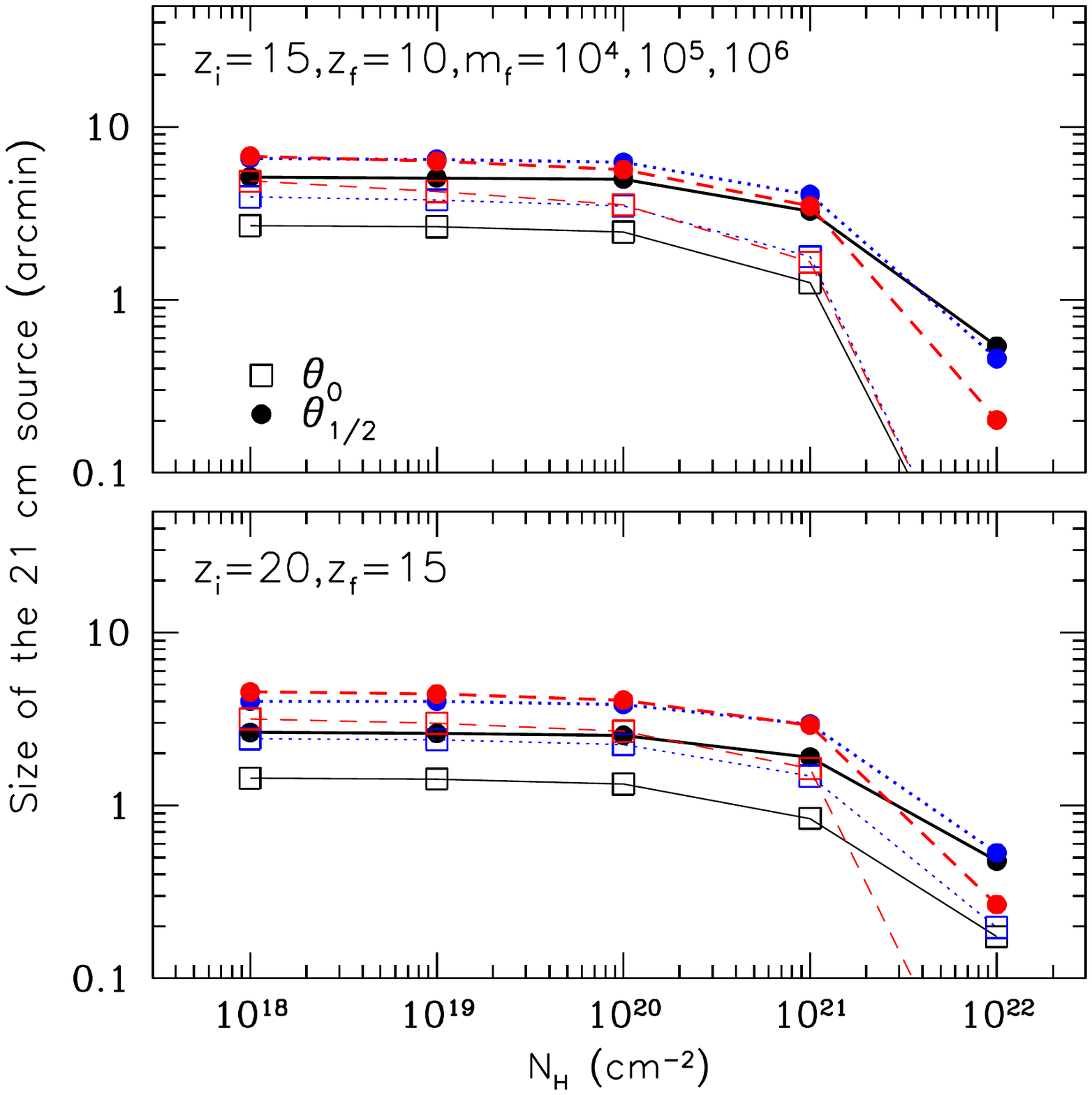}  
\caption{{\it Top panel:} Characteristic angular sizes (see text for
  definitions) $\rnull$ (empty squares connected by thin lines) and
  $\rhalf$ (filled circles connected by thick lines) of the 21~cm
  signal around a miniquasar as a function of the absorption column
  for $\zi=15$, $\zf=10$, $\mdot=1$ and three sets of BH masses
  ($\mi,\mf$): ($2\times 10^3$, $10^4$) -- solid black, ($2\times 10^4$,
  $10^5$) -- dotted blue, and ($2\times 10^5$, $10^6$) -- dashed
  red. {\it Bottom panel:} Similar, for $\zi=20$ and $\zf=15$.
}
\label{fig:radii}
\end{figure}

Figure~\ref{fig:radii} (top panel) shows $\rnull$ and $\rhalf$ as
functions of the absorption column for $\zi=15$, $\zf=10$, $\mdot=1$
and three different BH masses, $\mf=10^4$, $10^5$ and $10^6$. We see
that if the intragalactic absorption is not strong ($\NH\lesssim
10^{20}$~cm$^{-2}$), $\rnull$ and especially $\rhalf$ depend only
weakly on the BH mass and absorption column density. Specifically,
$\rnull\sim 2.5$--5~arcmin, and $\rhalf\sim 5$--7~arcmin. If
$\NH\gtrsim 10^{21}$~cm$^{-2}$, most of the microquasar's soft X-ray 
emission is absorbed within its host galaxy, which naturally leads to
a dramatic weakening of IGM heating and shrinkage of the 21~cm
zone. The bottom panel of Fig.~\ref{fig:radii} shows a similar
set of curves for the case of miniquasars operating at higher
redshifts, namely $\zi=20$ and $\zf=15$. In this case, there is a more
noticeable, albeit still weak dependence on the BH mass, namely (for
$\NH\lesssim 10^{20}$~cm$^{-2}$) $\rnull$ changes from $\sim 1.5$ to
$\sim 3$~arcmin as $\mf$ increases from $10^4$ to $10^6$, whereas
$\rhalf$ changes from $\sim 2.5$ to $\sim 4.5$~arcmin in the same BH
mass range. Overall, the computed $\rnull$ size of the heating zone is
in remarkably good agreement (within a factor of $\sim 2$) with our
rough prediction given by equation~(\ref{eq:path_arcmin}).

\begin{figure}
\centering
\includegraphics[width=\columnwidth,viewport=20 180 560 720]{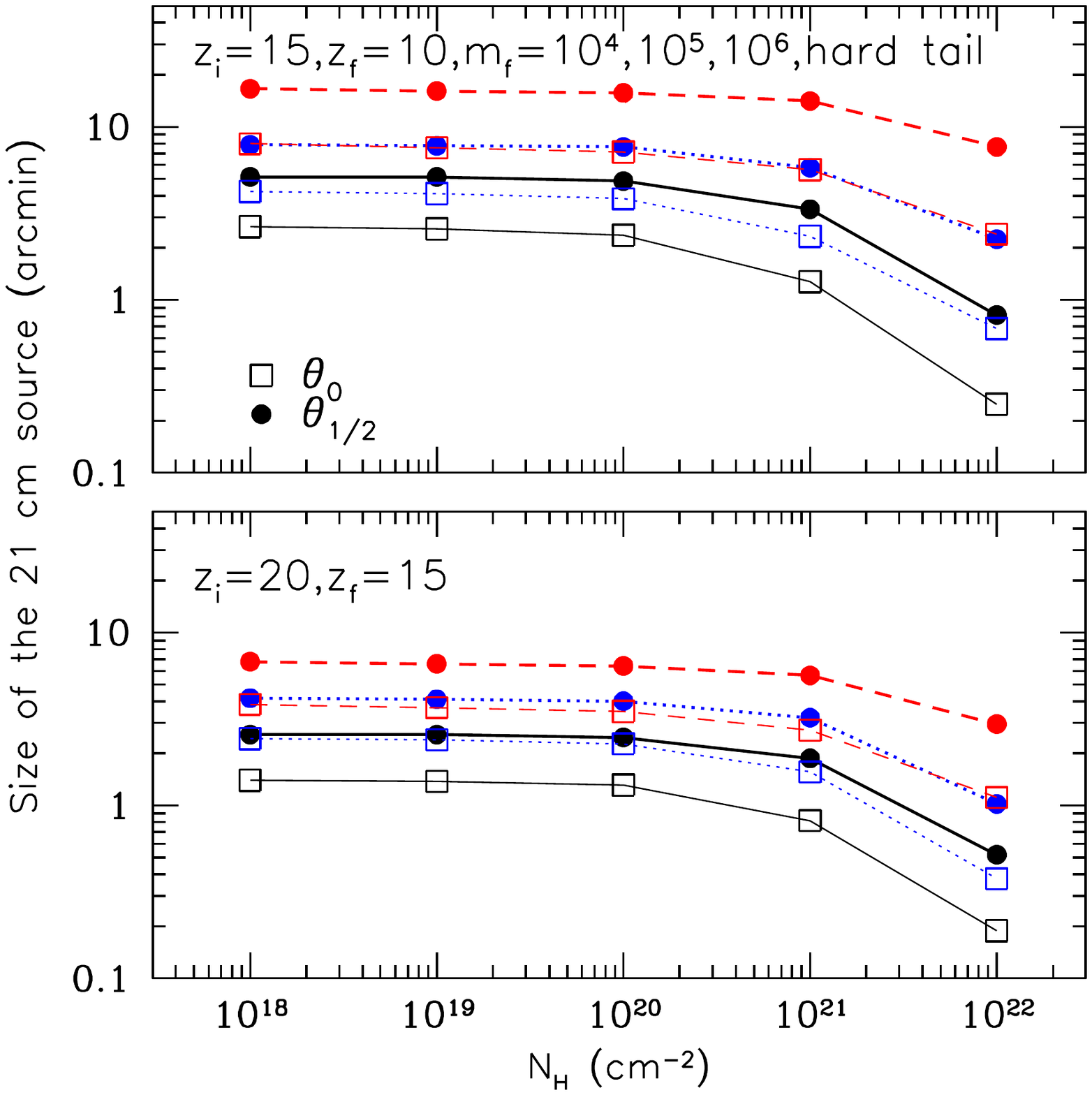}  
\caption{As Fig.~\ref{fig:radii}, but for thermal disk emission
  with a Comptonization tail ({\it simpl}*{\it diskbb}) instead of
  pure thermal disk emission.
}
\label{fig:radii_simpl}
\end{figure}

Figure~\ref{fig:radii_simpl} demonstrates the impact of an
additional hard (Comptonization) spectral component on the extent of
the miniquasar 21~cm zone. By comparing these plots with those
pertaining to the case of pure multicolor disk emission
(Fig.~\ref{fig:radii}), we see that while $\rnull$ and $\rhalf$ have
remained nearly unchanged for $\mf\le 10^5$, these characteristic
radii have increased by a factor of $\sim 1.5$--2 for
$\mf=10^6$. Therefore, the hard tail considerably changes the overall
picture for the most massive of the considered BHs ($m=10^6$). This
again reflects the fact that, within the adopted model, the
Comptonized radiation starts to dominate over the thermal emission
already at photon energies $\sim 300$~eV.

\subsection{Spectrum and flux of the 21~cm signal}
\label{s:spectrum21}

We now proceed to discussing the spectral properties of the 21~cm
signal associated with high-redshift miniquasars. Based on the above
results we can expect such objects to be surrounded on the sky by
fairly well defined regions with an apparent size of several arcmin
within which $\Tb-\Tbbgr\gtrsim 100$~mK, and it will be interesting to
search for such specific zones of 21~cm excess emission with future
radio interferometers.

In reality, the effective angular size of the 21~cm signal extraction
region around a candidate miniquasar will be determined by the
characteristics of a particular radio interferometer and by the
related noise and background levels (see the discussion in
\S\ref{s:summary} below), but ideally it should be of the order of the
$\rhalf$ radius defined above. We have therefore integrated the
surface brightness of the expected 21~cm excess emission (i.e the
difference $\Tb-\Tbbgr$) over the circle of radius $\rhalf$ around the
miniquasar.

\begin{figure}
\centering
\includegraphics[width=\columnwidth,viewport=20 180 560 720]{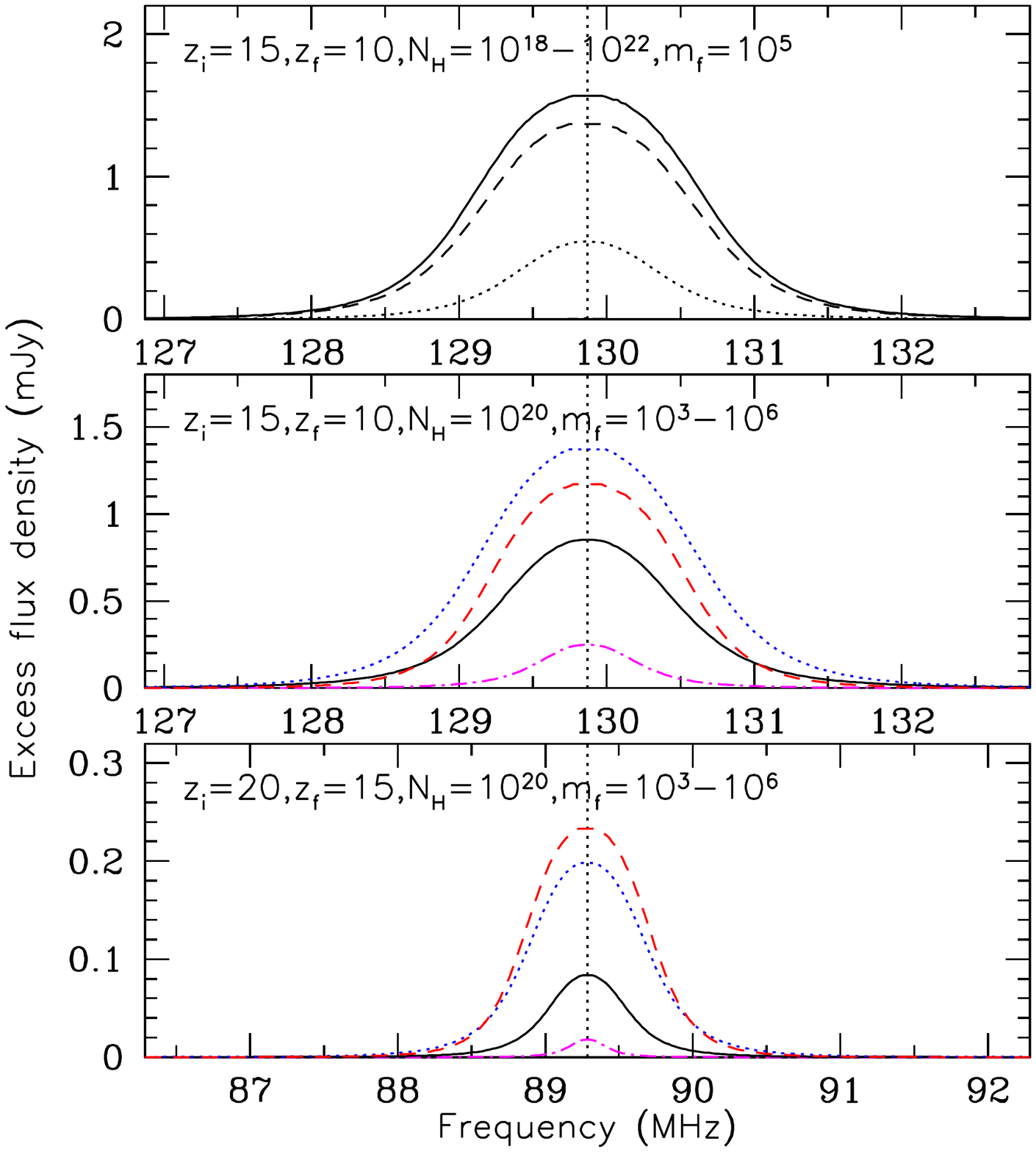}
\caption{{\it Top panel:} Spectra of 21~cm excess emission (with
  respect to the background level at the miniquasar's redshift)
  integrated within $\rhalf$ for $\zi=15$, $\zf=10$, $\mdot=1$,
  $\mi=2\times 10^4$, $\mf=10^5$ and various $\NH$ columns
  (cm$^{-2}$): $10^{18}$(solid), $10^{20}$ (dashed), $10^{21}$
  (dotted), $10^{22}$ (dash-dotted, poorly visible near the lower
  boundary of the plot). {\it Middle panel:} Similar, for a fixed
  absorption column ($\NH=10^{20}$~cm$^{-2}$) and various BH masses
  ($\mi$, $\mf$): ($2\times 10^2$, $10^3$) -- dash-dotted magenta,
  ($2\times 10^3$, $10^4$) -- solid black, ($2\times 10^4$, $10^5$) --
  dotted blue, and ($2\times 10^5$, $10^6$) -- dashed red. {\it Bottom
    panel:} Similar, for higher redshifts, $\zi=20$,
  $\zf=15$.
}
\label{fig:spectrum21}
\end{figure}

Figure~\ref{fig:spectrum21} (top panel) shows the resulting spectra
for $\zi=15$, $\zf=10$, $\mi=2\times 10^4$, $\mf=10^5$, $\mdot=1$ and
various absorption columns. We see that the 21~cm flux density is
almost unaffected by intragalactic absorption if $\NH\lesssim
10^{20}$~cm$^{-2}$, the signal weakens by a factor of $\sim 3$ for
$\NH=10^{21}$~cm$^{-2}$ and nearly vanishes if
$\NH=10^{22}$~cm$^{-2}$, as essentially no soft X-rays from the
miniquasar leak from the host galaxy into the IGM.

The middle panel of Fig.~\ref{fig:spectrum21} demonstrates the
dependence of the 21~cm spectum on the BH mass (for $\zi=15$,
$\zf=10$, $\mdot=1$ and $\NH=10^{20}$~cm$^{-2}$). We see that the
signal increases by a factor of $\sim 3$ on going from $\mf=10^3$ to
$10^4$ and then remains nearly the same (within a factor of $\sim
1.5$) for $\mf=10^4$--$10^6$. The bottom panel of
Fig.~\ref{fig:spectrum21} shows the corresponding spectra for similar
miniquasars at higher redshifts: $\zi=20$, $\zf=15$. The picture is
qualitatively similar to the previous case, but the 21~cm excess flux
density is more sensitive to the BH mass at the higher redshift. We
also note that the signal is somewhat stronger for $\mf=10^5$ than for
$\mf=10^6$ in the $\zf=10$ case, while the opposite is true for
$\zf=15$. The reason is that this signal is accumulated from the
$\rhalf$ region whose dependence on the BH mass (see
Fig.~\ref{fig:radii}) is slightly different between $\zf=10$ and
$\zf=15$ due to a non-trivial interplay between the BH soft X-ray
spectral properties and redshift-dependent density of the IGM. Most
importantly, however, Fig.~\ref{fig:spectrum21} demonstrates that the
expected 21~cm signal depends fairly weakly (within a factor of 3) on
the BH mass over the $10^4$--$10^6\Msun$ range.

\begin{figure}
\centering
\includegraphics[width=\columnwidth,viewport=20 180 560 720]{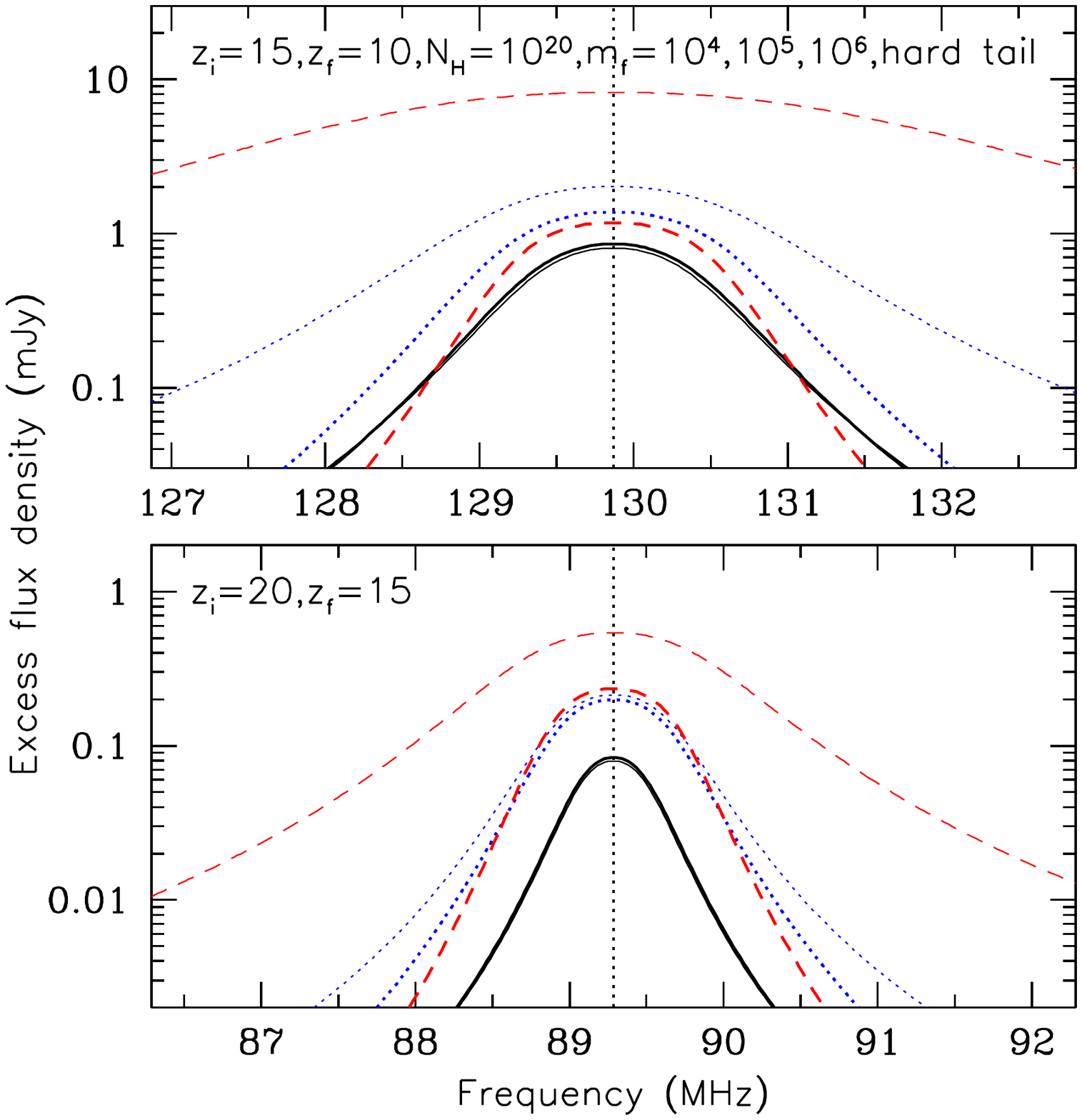}
\caption{Similar to Fig.~\ref{fig:spectrum21}, comparing the case of
  pure thermal disk emission ({\it diskbb}, thick curves) with that of
  a combination of disk emission and a Comptonization tail ({\it
    simpl}*{\it diskbb}, thin curves), for three sets of BH masses
  ($\mi$, $\mf$): ($2\times 10^3$, $10^4$) -- solid black, ($2\times 10^4$,
  $10^5$) -- dotted blue and ($2\times 10^5$, $10^6$) -- dashed red. Note the
  logarithmic vertical scale.
}
\label{fig:spectrum21_simpl}
\end{figure}

Figure~\ref{fig:spectrum21_simpl} demonstrates the impact of an
additional hard spectral component on the discussed 21~cm spectra. We 
see that the hard tail leads to a dramatic increase of the expected
21~cm signal for our most massive ($\mf=10^6$) BH, with this difference
being more pronounced at the lower redshift ($\zf=10$
vs. $\zf=15$). These tendencies are expected, since the reported
spectra were obtained by integration of $\Tb$ within the
characteristic radius $\rhalf$, which increases in the presence of a
hard spectral component, as was shown in \S\ref{s:radii}.

As regards the absolute value of the expected 21~cm flux density, it
is useful to approximate it as follows:
\beqa
\Fnu\approx \frac{2k}{(21~{\rm
    cm})^2(1+z)^2}\frac{3|\Tbbgr|}{4}\pi\rhalf^2,\nonumber\\
\approx
0.6\left(\frac{1+\zf}{11}\right)^{-2}\frac{|\Tbbgr|}{250~{\rm
    mK}}\left(\frac{\rhalf}{5'}\right)^2~{\rm mJy},
\label{eq:fnu}
\eeqa
where we assumed that the average excess brightness temperature of the
21~cm signal within $\rhalf$ is $3\Tbbgr/4$, which is approximately
the case (see the $\Tb$ radial profiles in
\S\ref{s:profiles}). Substituting the typical values derived from our
simulations for $\zf=10$ ($\Tbbgr=-309$~mK, $\rhalf=6'$) and $\zf=15$
($\Tbbgr=-245$~mK, $\rhalf=4'$) into the above expression, we find
$\Fnu\approx 1.1$ and $0.18$~mJy, respectively, in fairly good
agreement with the 21~cm spectra shown above.

We finally note that the simulated 21~cm spectra for the case of
purely thermal disk emission are characterized by FWHM of $\approx
0.01$ in terms of $\Delta\nu/\nu$.

\section{Relation to X-ray observations}
\label{s:xray}

As was discussed in \S\ref{s:intro}, there is a hope that future X-ray
observatories such as {\it Lynx} will be able to find a significant
number of high-redshift miniquasar candidates. Provided that the
planned next-generation radio facilities such as SKA are also
available by that time, it should be possible to search for the
specific 21~cm signatures of X-ray selected miniquasars discussed in
this paper. 

The first practical question then is: what are the limiting BH mass
and redshift for X-ray detection of miniquasars? In this study, we
have assumed that a miniquasar's spectrum is a combination of (i)
multicolor disk emission with the temperature expected from the
standard accretion disk theory and, plausibly, (ii) a hard, power-law
($\Gamma\approx 2$) tail associated with Comptonization of disk
emission in a hot corona, which extends to high energies (at least to
a few tens of keV). It is this additional hard component that an X-ray
telescope might be able to detect from a high-redshift miniquasar. In
the particular spectral model, {\it simpl(diskbb)}, used in this
study, the observed X-ray flux in the 0.5--2~keV band (corresponding
to emission in a rest-frame band of $0.5(1+z)$--$2(1+z)$~keV), $\Fx$,
is proportional to the fraction of scattered photons. It turns out
that, virtually independently of the BH mass and accretion rate, 
\beq
\Fx\approx 0.06\frac{\fsc}{0.05}\frac{L}{4\pi\Dl^2},
\label{eq:fx}
\eeq
where $L$ is the total luminosity of the miniquasar and $\Dl(z)$ is
its luminosity distance.

Assuming that $\fsc=0.05$ (which is a reasonable value as discussed in
\S\ref{s:abs}) and that the X-ray tesecope catches the miniquasar when
it is accreting at a critical rate ($\mdot=1$), when $L=\Le(m)$, we
find from equation~(\ref{eq:fx}) that for $m=10^4$, $\Fx=2.3\times
10^{-20}$ and $5.8\times 10^{-20}$~erg~cm$^{-2}$~s$^{-1}$ at $z=15$
and $z=10$, respectively, whereas for $m=10^5$, $\Fx=2.3\times
10^{-19}$ and $5.8\times 10^{-19}$~erg~cm$^{-2}$~s$^{-1}$ for the same
redshifts. These fluxes are well below the detection threshold of {\it
  Chandra}, the most sensitive X-ray telescope so far. However, the
proposed {\it Lynx} mission is expected to reach a sensitivity of
$\sim 10^{-19}$~erg~cm$^{-2}$~s$^{-1}$ in its deep extragalatic
surveys and should thus be able to detect actively growing BHs of mass
$\gtrsim 5\times 10^4\Msun$ at $z=15$ and $\gtrsim 2\times 10^4\Msun$
at $z=10$. These mass limits are of course inversely proportional to
$\mdot$ and $\fsc$. 

\section{Discussion and summary}
\label{s:summary}

We have shown that an intermediate mass BH growing by radiatively
efficient accretion in the early Universe should leave a specific,
localized imprint on the 21~cm cosmological signal. Namely, a
miniquasar with the BH mass between $\sim 10^4$ and $\sim 10^6\Msun$ 
at $z\sim 15$--10 will be surrounded by a region with a fairly well
defined boundary of several arcmin radius, within which the 21~cm
temperature quickly grows inwards from the background value
$\Tbbgr\sim -250$~mK to $\Tb\sim 0$ (reaching the saturation value of 
$\sim 30$~mK in the innermost region). The size of this region and the
flux density of the enclosed 21~cm signal are only weakly sensitive to
the BH mass in the range quoted above.

\subsection{Sensitivity to assumptions}
\label{s:assumptions}

The above result was obtained under certain assumptions and it is
important to discuss how realistic they are. Perhaps, the most
important constituent of our model is the miniquasar's spectral energy
distribution, which we assumed to be that of multicolor disk blackbody
emission. As was discussed in \S\ref{s:spec}, the actual spectrum of the
accretion disk emission is likely to deviate significantly from this
simplistic model, but given the weak sensitivity of the size of
the 21~cm zone to the BH mass, such deviations are unlikely to have a
significant effect on our predictions. More important is the likely
presence of an additional, hard component in the miniquasar
spectrum. As we have demonstrated, its effect is small for relatively
low-mass BHs ($\sim 10^4$--$10^5\Msun$) but becomes substantial (the
heating zone widens by a factor of $\sim 1.5$--2) for a $10^6\Msun$ BH.

The next important issue is possible photoabsorption of the
miniquasar's soft X-ray emission within its host galaxy. It turns out
that the properties of the 21~cm zone remain nearly unchanged as long
as $\NH\lesssim 10^{20}$~cm$^{-2}$  (regardless of the presence of
metals in the absorbing medium), but at $\NH\sim 10^{21}$~cm$^{-2}$
this zone starts to shrink dramatically. One may argue that a powerful
miniquasar should be able to quickly photoionize 
the interstellar medium within a substantial distance of itself and
thus effectively reduce $\NH$ (e.g. \citealt{sazkha18}), but this
clearly needs further investigation. Furthermore, if miniquasars are
less powerful analogs of AGN, they may have a small-scale obscuring
torus of cold gas and dust. In that case there will be two opposite
cones of specific 21~cm signal around the miniquasar, i.e. the average
signal within the $\rhalf$ radius will decrease by a factor of
$\Omega/4\pi$, where $\Omega$ is the solid angle of the unobscured sky
as seen from the BH.

Finally, we assumed that the Universe had not yet been globally heated
at the redshifts of interest ($z\sim 15$--10) and that the 21~cm spin
temperature was coupled to the gas temperature at these epochs. It is
only in this case that a large contrast in the 21~cm brightness
temperature will arise between the vicinity of the miniquasar, where
$\Tb\sim 0$, and the background, where $\Tb\sim -(200$--$300)$~mK (at
$z=15-10$). These assumed conditions are in good
agreement with the recent EDGES result \citep{bowetal18} for $z\sim
20$--15\footnote{Actually, EDGES measured an even lower sky-averaged
  $\Tb\sim -500$~mK.} but appear to fail at $z\lesssim 14$, when the
global 21~cm temperature has been measured to be around zero. Of course, the
EDGES measurements must be verified with future observations but a lot
of authors (see \S\ref{s:intro}) have suggested that XRBs, miniquasars
and other types of X-ray sources can indeed significantly heat the
Universe by $z\sim 10$. In such a case, it will be extremely difficult
to discern the 21~cm imprint of an individual miniquasar against the
background at $z\sim 10$ but that should still be possible at
$z\sim 15$.

\subsection{Comparison with previous studies}
\label{s:previous}

The present study is not the first one addressing the potential impact
of high-redshift X-ray sources on the cosmological 21~cm signal. In
particular, a number of authors have focused on the expected 21~cm
signatures of individual quasars and miniquasars in the early Universe
\citep{chuetal06,thozar08,yajli14,ghaetal17,boletal18,vasetal18}. The
crucial novel aspect of our study is its focus on the (relatively
soft) thermal emission of the accretion disk around an
intermediate-mass black hole, which is expected (based on the rich
observational material on high Eddington ratio X-ray binaries and AGN)
to carry the bulk of the bolometric luminosity of the miniquasar but
has been usually ignored before. This leads to an important difference
for the predicted 21~cm signature, namely that it should be
concentrated within $\sim 5$~arcmin of the miniquasar due to the
relatively short mean free path of the extreme UV/soft X-ray photons
in the ambient IGM.

Furthemore, in contrast to some of the previous studies we have
assumed the 21~cm spin temperature at the considered epochs ($z\sim
15$--10) to be coupled (by the UV background from the first stars) to
the gas temperature throughout the IGM and that the latter had not yet
been heated significantly on average, so that the mean $\Tb\sim
-(200$--$300)$~mK, rather than $\Tb\approx 0$ as it would be in the
absence of Wouthuysen--Field coupling or in the presence of
significant global heating. This key assumption, as noted above (in
\S\ref{s:assumptions}), is partially motivated by the recent detection
of a strong 21~cm global absorption feature by EDGES. It is the
combination of the relative compactness of the heating zone and the
large negative global 21~cm brightness temperature that has led to our
conclusion that high-redshift miniquasars might be associated with
fairly strong ($\sim 0.2[(1+z)/16]^{-2}$~mJy) 21~cm signatures.

\subsection{Observational strategy}
\label{s:strategy}

A blind sky search for weak 21~cm signals from individual
high-redshift miniquasars might not be feasible in the near
future. Therefore, we propose to look for such signals specifically
from miniquasar candidates that can be found with the proposed {\it
  Lynx} X-ray mission. As discussed in \S\ref{s:xray}, the planned
{\it Lynx} sensitivity of $\sim 10^{-19}$~erg~cm$^{-2}$~s$^{-1}$
should allow it to detect rapidly growing BHs with masses as low as a
few $10^4\Msun$ out to $z\sim 15$, provided that a signficant fraction
of the energy released by accretion goes into Comptonized, hard X-ray
radiation.

However, selection of such candidates is unlikely to be an easy
task. Indeed, {\it Lynx} will provide only crude X-ray hardness
information for them, not sufficient for distinguishing from other
types of sources. Moreover, high-redshift miniquasars will probably be
just a small minority among the tens of thousands of sources to be
detected in the proposed {\it Lynx} (400~arcmin$^{2}$) ultradeep
survey \citep{lynx18}. Specifically, \cite{benetal18} predict that
between several dozen and a few thousand growing massive BHs (roughly
in the $10^4$--$10^6\Msun$ range of interest to us) could be found at
$z\sim 5$--12, with this large uncertainty reflecting our poor
understanding of how BHs form and grow in the early Universe. In the
present study, we have focused on somewhat earlier epochs, $z\sim
15$--10, and just between a few and a few hundred (i.e. less or much
less than 1 object per arcmin$^2$) such high-redshift miniquasars are
expected to be found by {\it Lynx} \citep{benetal18}.

Fortunately, there are bright prospects for the synergy between the
proposed {\it Lynx} survey and the optical/IR ultradeep surveys by
next-generation telescopes such as {\it JWST} \citep{lynx18}, so that
the majority of the {\it Lynx} sources (such as AGN and galaxies) will
probably have reliable optical/IR counterparts. High-redshift
miniquasars, because of their expected optical/IR faintness (see
\S\ref{s:intro}), will thus be hidden among the relatively small
sample of very faint ($\Fx\gtrsim 10^{-19}$~erg~cm$^{-2}$~s$^{-1}$)
X-ray sources {\it without} an optical counterpart. A reasonable
approach would then be to regard all such sources as candidate
high-redshift miniquasars and search for specific 21-cm signatures
around their positions provided by {\it Lynx}. 

\begin{figure}
\centering
\includegraphics[width=\columnwidth,viewport=20 180 560 720]{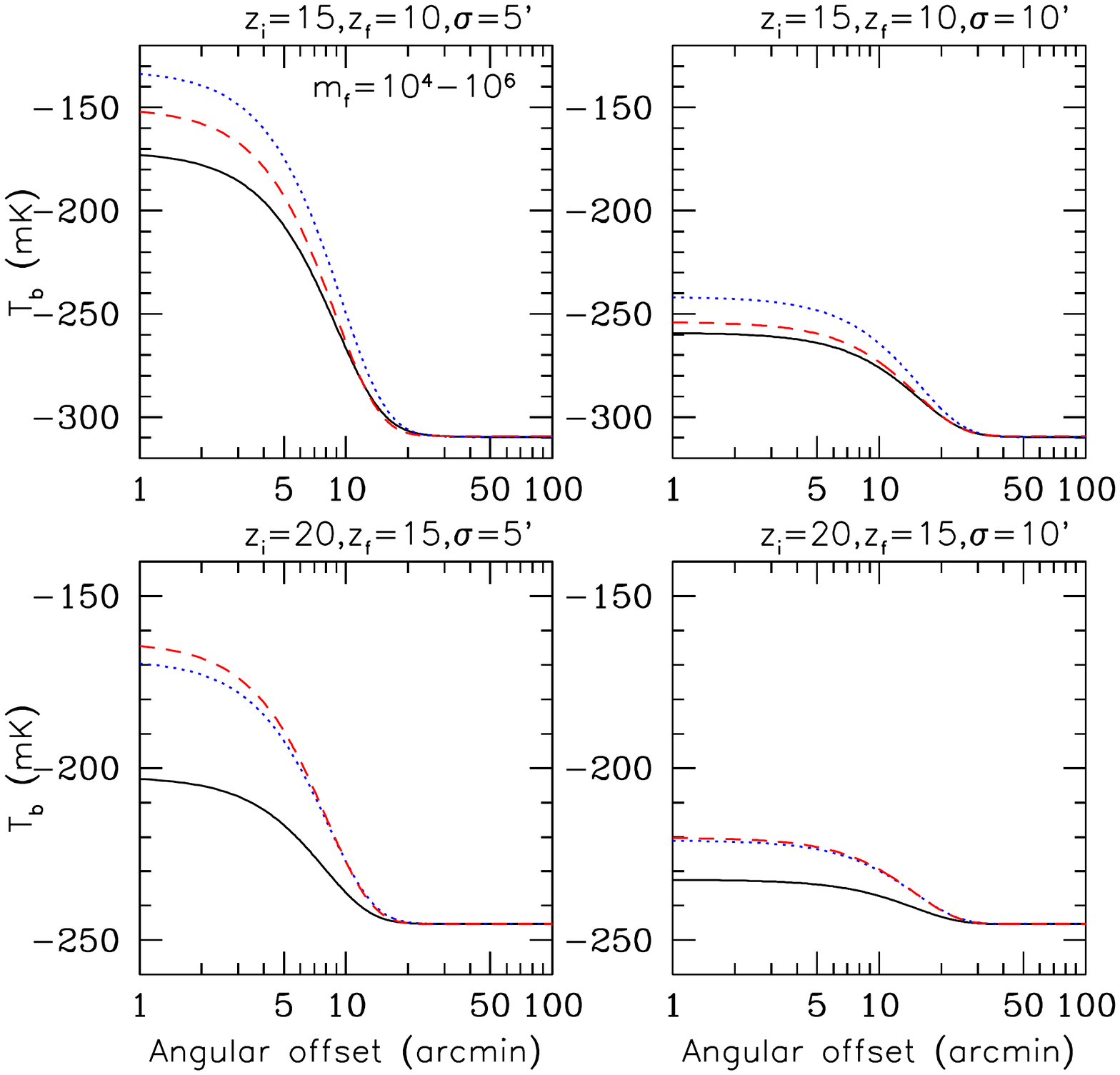}
\caption{Radial profiles of the 21~cm images of miniquasars after
  convolution with a 2-dimensional Gaussian with $\sigma=5'$ (left)
  and $10'$ (right). The upper panels are for $\zi=15$ and $\zf=10$,
  and the lower ones for $\zi=20$ and $\zf=15$. The different curves
  correspond to different BH masses ($\mi$, $\mf$): ($2\times 10^3$,
  $10^4$) -- solid black, ($2\times 10^4$, $10^5$) -- dotted blue,
  ($2\times 10^5$, $10^6$) -- dashed red. These results are for
  multicolor disk emission and $\NH=10^{20}$~cm$^{-2}$. Note the
  logarithmic horizontal scale.
}
\label{fig:profiles_gauss}
\end{figure}

According to our estimates, high-redshift miniquasars are expected to
produce 21~cm signals with an amplitude $\sim 100$~mK on few-arcmin
scales, with the characteristic spectral width $\Delta\nu/\nu\sim
0.01$ corresponding to $\Delta\nu\sim 1$~MHz at $\nu\sim 100$~MHz for
$z=10$--15. How do these numbers compare with the expected
characteristics of future cosmological 21~cm surveys with their
specific noise levels and angular resolution?
Figure~\ref{fig:profiles_gauss} shows the result of convolution of our
predicted 21~cm images of miniquasars (for $m=10^4$-$10^6$ and
$\zf=10$ and 15) with two-dimensional Gaussians with $\sigma=5'$ and
10$'$. We see that miniquasars are expected to produce $\sim
140$--180~mK and $\sim 50$--80~mK positive peaks (with respect to the
large-scale background) on images with 5-arcmin angular resolution for
$z\sim 10$ and $z\sim 15$, respectively. 

The low-frequency component of the SKA experiment, SKA-low, is planned
to cover a broad frequency range extending down to $\sim 50$~MHz
(allowing one to probe the early Universe out to $z\sim 25$) with high
spectral resolution ($\sim 1$~kHz) and a very large collecting area
$\sim 1$~km$^2$ \citep{meletal13}. For the first phase of the
experiment (the so-called SKA1-Low), the noise level is expected to be
$\sim 3$~mK ($\sim 10$~mK) for $z=10$ ($z=15$) images with $5'$
angular resolution (with the frequency bandwidth matched to the
angular resolution) accumulated over an integration time of $\sim
1000$~hours (see fig.~2 in \citealt{meletal15}). Therefore, the $\sim
100$~mK signal from a high-redshift miniquasar (see
Fig.~\ref{fig:profiles_gauss}) should be reliably detectable with such
long (but feasible) observations by SKA1-low (and even more so by the
fully constructed SKA-low).

In reality, the biggest problem will likely be separating the 21~cm
signal associated with high-redshift miniquasars from astrophysical
foregrounds of much higher amplitude, such as Galactic synchrotron
radiation and cumulative emission from unresolved extragalactic
sources; furthermore, the large-scale structure of the early Universe
will cause additional fluctuations of the 21~cm brightness temperature
on the arcmin scales relevant to the problem in hand (see
\citealt{meletal13} for a review). Finally, as with any radio
interferometer, SKA will not be capable of measuring absolute source
fluxes due to the zero-spacing problem. A thorough consideration of
these non-trivial observational issues is beyond the scope of this
proof-of-concept study, but a number of recent studies
\citep{wyietal15,ghaetal17} address these problems in the context of
SKA and suggest that there are efficient methods of evaluation and
subtraction of foregrounds that might enable detection of the $\sim
100$~mK peaks associated with high-redshift miniquasars on the SKA
images.

We finally emphasize again that such a search will be greatly
faciliated by the availability of accurate celestial coordinates of
candidate miniquasars from {\it Lynx}, even though the X-ray data will
not provide their redshifts. In practice, the search might consist of
browsing SKA-low images constructed with $\sim 5'$ angular resolution
(and cleaned as carefully as possible from the foregrounds), one
redshift slice after another over a range of $z\sim 10$--20. The
detection of a positive $\sim 100$~mK peak (see
Fig.~\ref{fig:profiles_gauss}) centered on the {\it Lynx} position
will strongly indicate that the object is indeed an intermediate-mass
BH growing to become a supermassive BH. Moreover, its redshift can
thus be measured to within $\Delta z/(1+z)\lesssim 0.01$.

\section*{Acknowledgments}

The authors thank the referee for useful suggestions. The research was
supported by the Russian Science Foundation (grant 14-12-01315).
  

\bsp	
\label{lastpage}
\end{document}